\documentstyle[pre,aps,epsf,fancyhdr]{revtex}

\begin{document}

\draft

\title
{Generalized empty-interval method applied to a class of one-dimensional stochastic models\footnote{To appear in Physical Review E (November 2001)}}

\author
{Mauro Mobilia and Pierre-Antoine Bares  }

\address
{Institute of Theoretical Physics, Swiss Federal Institute of Technology of  Lausanne, CH-1015 Lausanne  EPFL, Switzerland }

\date{\today} 

\maketitle


\begin{abstract}

In this work we study, on a finite and periodic lattice, a class of one-dimensional (bimolecular and single-species) reaction-diffusion models which cannot be mapped onto free-fermion models. We extend the conventional empty-interval method, also called  {\it interparticle distribution function} (IPDF) method, by introducing a string function, which is simply related to relevant physical quantities.

 As an illustration, we specifically consider a model which cannot be solved directly by the conventional IPDF method and which can be viewed as a generalization of the {\it voter} model and/or as an {\it epidemic} model. We also consider the {\it reversible} diffusion-coagulation model with input of particles and determine other reaction-diffusion models which can be mapped onto the latter via suitable {\it similarity transformations}. Finally we study the problem of the propagation of a wave-front from an inhomogeneous initial configuration and note that the mean-field scenario predicted by Fisher's equation is not valid for the one-dimensional (microscopic) models under consideration.
  
\end{abstract}

\pacs{PACS number(s): 02.50.-r, 02.50.Ey, 05.50.+q, 82.40.-g}
\section{Introduction}

Reaction-diffusion models (RD) play an important role  in the description of classical interacting many-particle non-equilibrium systems and have been extensively  investigated in the last decade \cite{Privman,Schutzrev}.
Often these systems have been treated by mean-field techniques (e.g. rate equations) which give rise to non-linear partial differential equations (such as, e.g., the Fisher equation \cite{Fisher}). The latter represent difficult mathematical problems: e.g. the Fisher equation cannot, in general, be solved exactly. The mean-field methods can accurately describe the behaviour of RD systems in {\it higher dimensions}, where the correlations do not dramatically change the physics of the models. However, in one spatial dimension where the fluctuations play a crucial role, these mean-field treatments fail. 
 In this sense, a satisfactory understanding of RD models in lower dimensions requires  {\it exact solutions}, which are scarce. In some cases, however, some RD models are known to be solvable. These cases can essentially be classified into four categories:
(i) models for which the equations of motion of correlation functions are closed \cite{Schutz}; (ii) the {\it free-fermion} models \cite{free-fermions} (or systems which can be mapped onto the latter, see \cite{Schutzrev,Henkel}); some other one-dimensional RD models can be solved by the (iii) {\it Matrix Ansatz} method \cite{Derrida} first introduced to study the steady-states of the {\it asymmetric exclusion process} (ASEP) and which has been extended to other multispecies RD models where the total number of particles is conserved. A dynamical version of the {\it matrix Ansatz} 
\cite{DMA} has also been proposed to study the dynamical properties of the models for which the
 equations are closed (on periodic as well as open chains). Some other one-dimensional models can be solved  by (iv) the empty-interval method, also called {\it interparticle distribution function} (IPDF) method \cite{Doering,benA,Hinrichsen,Peschel}, first introduced for the study of the diffusion-coagulation model. The solution of various one-dimensional RD models have been obtained from the diffusion-coagulation model via {\it similarity transformations}\cite{Schutzrev,Simon}. It has been established that the latter solvable models correspond to {\it free-fermion} systems \cite{Schutzrev}.

The purpose of this work is to present a generalization of the IPDF method and to apply this technique to solve a class of one-dimensional stochastic models which cannot be mapped onto free-fermion systems. In fact much attention has been given to {\it free-fermion} systems in various contexts (using {\it fermionic} algebra \cite{free-fermions} or via the {\it traditional} IPDF method, in the continuum limit \cite{Doering,benA} as well as on discrete lattice \cite{Hinrichsen,Simon}). The situation is different for the models considered here, for which only a few exact results are known.   

The paper is organized as follows: in the next section we briefly recall the formalism which we employ. In section III we introduce the string-function which is the key to our analysis and determine the constraints necessary to have solvable situations. In section IV we solve the general equations of motion for the string-functions of reaction-diffusion models which cannot be mapped onto free-fermion systems. The latter provides the exact expression of the density and the instantaneous nearest-neighbour (two-point) correlation function. We also present an approximative,
 and, recursive scheme to compute the (other) instantaneous two-point correlation functions. In section V, we specifically consider a model with branching and coagulation reactions which cannot be solved  directly by the traditional IPDF method. In section VI, we solve a reversible  diffusion-coagulation model with external input of particles. In section VII, we take advantage of the results of the previous section to solve other related models via similarity transformations. 
In section VIII we study, for the models introduced and solved in section V and VI, the problem of the propagation of a wave-front starting from an inhomogeneous initial state and observe that the {\it mean-field} scenario predicted by {\it Fisher's equation}\cite{Fisher,Riordan,Murray} is not valid at the microscopic level. Finally, the section IX is devoted to the conclusion.

\section{The Formalism}

Before generalizing the IPDF method, it is useful to briefly review the so-called {\it stochastic Hamiltonian} formalism . 

It is known that models  of stochastic hard-core particles are soluble on some manifold on wich the equations of motion of their correlation functions close \cite{Schutz}. In this work, we concentrate on one-dimensional bimolecular single-species reaction-diffusion systems.

Consider a periodic chain   with $L$ sites (labelled from $1$ to $L$).
On the lattice, local bimolecular reactions between single-species particle  ${\cal A}$, with hard-core, take place. Each site can be empty (denoted by the symbol $0$)  
or occupied at most by a particle of type ${\cal A}$
denoted in the following by the
index $1$ .
The reactions occuring on the sites $j$ and $j+1$ are specified by the transition rates, which here are assumed to be {\it site-dependent}, according to
$\Gamma_{\alpha \beta}^{\gamma \delta}$, where $\alpha,\beta,\gamma,
\delta=0,1$: $\forall (\alpha, \beta)\neq (\gamma, \delta),  \Gamma_{\alpha \beta}^{\gamma \delta}: \alpha + \beta \longrightarrow \gamma +\delta $.
Probability conservation implies $\Gamma_{\alpha \beta}^{\alpha \beta}=-\sum_{(\alpha,\beta)\neq(\alpha',\beta')}
\Gamma_{\alpha \beta}^{\alpha' \beta'}$
and $\Gamma_{\alpha \beta}^{\gamma \delta}\geq 0, \; \forall (\alpha, \beta)\neq (\gamma, \delta) $

For example the rate $\Gamma_{1 1}^{1 0}$ corresponds to the reaction
${\cal A} {\cal A} \longrightarrow {\cal A} \emptyset$ and  $\Gamma_{1 1}^{1 1}=-(\Gamma^{1 0}_{1 1}+\Gamma^{0 1}_{1 1}+\Gamma^{0 0}_{1 1}).$

The state of the system is represented by the ket $|P(t)\rangle=\sum_{\{n\}}P(\{n\},t)|n \rangle$,
where the sum runs over the $2^{L}$ configurations.
At site $i$ the local state is specified by the ket 
$|n_i\rangle =(1 \; 0 )^{T}$ if the site 
$i$ is empty,  $|n_i\rangle =(0 \; 1 )^{T}$ if the site $i$ is occupied by 
a particle of type ${\cal A}$ 
($1$) .

It is by now well established that a master equation can be rewritten formally as an imaginary time Schr\"odinger equation: 
$ \frac{\partial}{\partial t}|P(t)\rangle = -H |P(t)\rangle,$
where $H$ is the {\it Stochastic Hamiltonian} which governs the dynamics of the system. In general, it is neither hermitian nor normal. Its construction from the master equation is a standard procedure (see e.g. \cite{Privman,Schutzrev})
The evolution operator $H=\sum_{j=1}^{L}H_{j,j+1}$ acts locally on two adjacent sites, with  
$
-H_{j, j+1} =
\left(
 \begin{array}{c c c c}
 \Gamma_{0 0}^{0 0} & \Gamma_{0 1}^{0 0} &\Gamma_{1 0}^{0 0} &\Gamma_{1 1}^{0 0}  \\
 \Gamma_{0 0}^{0 1} & \Gamma_{0 1}^{0 1} &\Gamma_{1 0}^{0 1} &\Gamma_{1 1}^{0 1} \\
 \Gamma_{0 0}^{1 0} & \Gamma_{0 1}^{1 0} &\Gamma_{1 0}^{1 0} &\Gamma_{1 1}^{1 0} \\
 \Gamma_{0 0}^{1 1} & \Gamma_{0 1}^{1 1}&\Gamma_{1 0}^{1 1} &\Gamma_{1 1}^{1 1} \\
 \end{array}\right)\ ,
$
where the same notations as in reference \cite{Schutz,Fujii,Mobar2} have been used.
Probability conservation implies that each column in the above representation
sums up to zero. 

The {\it left vacuum} $\langle \widetilde \chi|$,  which is defined as  $\langle \widetilde \chi|\equiv \sum_{\{n\}} \langle\{n\} |$, locally   has the representation
$\langle \widetilde \chi |=(1\;  1)\otimes (1 \;  1)$ with the property:
$\langle \widetilde \chi|H_{j, 
j+1}=0$.

Below we shall assume an initial state $|P(0)\rangle$ and investigate the
expectation value of an operator $O$ (observables such as density etc.): 
$
\langle O \rangle(t)\equiv \langle \widetilde \chi|O e^{-Ht}|P(0)\rangle
$. 
For general $s-$species bimolecular reaction-diffusion systems, there are $(s+1)^4$ possible rates which have to fulfill the $(s+1)^2$  probability conservation constraints. Thus general $s-species$ bimolecular reaction-diffusion systems are characterized by $(s+1)^4-(s+1)^2$ independent parameters \cite{Fujii,Mobar2}.  If one imposes to these
 parameters $2s^3$ appropriate constraints, the equation of motion of correlation functions close and the system is  formally soluble in arbitrary dimensions. Here, we focus on the case $s=1$, and so we have $16-4=12$ independent rates and $2$ {\it closure constraints}.

For single-species bimolecular processes, 
with the notations:
\begin{eqnarray}
\label{eq.0.9}
A_{1}&\equiv& \Gamma_{0 0}^{0 1}+\Gamma_{0 0}^{1 1}\;,\;
 B_{1}\equiv  \Gamma_{1 0}^{0 1}+\Gamma_{1 0}^{1 1}- \Gamma_{0 0}^{0 1}-\Gamma_{0 0}^{1 1}\;,\; 
C_{1}\equiv  \Gamma_{0 1}^{0 0}+\Gamma_{0 1}^{1 0} +\Gamma_{0 0}^{0 1}+\Gamma_{0 0}^{1 1}\;,
D_{1}\equiv C_{1}-(\Gamma_{1 0}^{0 1}+\Gamma_{1 0}^{1 1}+\Gamma_{1 1}^{0 0}+\Gamma_{1 1}^{1 0})\;, \nonumber\\
A_{2}&\equiv& \Gamma_{0 0}^{1 0}+\Gamma_{0 0}^{1 1}\;,\;
 B_{2}\equiv  \Gamma_{0 1}^{1 0}+\Gamma^{1 1}_{0 1}- \Gamma_{0 0}^{1 0}-\Gamma_{0 0}^{1 1}\;,\; 
C_{2}\equiv  \Gamma_{1 0}^{0 0}+\Gamma_{1 0}^{0 1} +
\Gamma_{0 0}^{1 0}+\Gamma_{0 0}^{1 1}\;, 
D_{2}\equiv C_{2}-(\Gamma_{0 1}^{1 0}+\Gamma_{0 1}^{1 1}+\Gamma_{1 1}^{0 0}+\Gamma_{1 1}^{0 1})\;, 
\end{eqnarray}
the  {\it closure constraints} are the following \cite{Schutz}:
\begin{eqnarray}
\label{eq.0.7}
D_{2}&=&0\Rightarrow \Gamma_{0 0}^{1 0}+ \Gamma_{0 0}^{1 1}-\left(\Gamma_{1 1}^{0 0}+\Gamma_{1 1}^{0 1}  \right)= \Gamma_{0 1}^{1 0} + \Gamma_{0 1}^{1 1}-\left(\Gamma_{1 0}^{0 0}+\Gamma_{1 0}^{0 1}  \right)\;,\nonumber\\
D_{1}&=&0\Rightarrow \Gamma_{0 0}^{0 1} + \Gamma_{0 0}^{1 1}-\left(\Gamma_{1 1}^{0 0}+\Gamma_{1 1}^{1 0}  \right)= \Gamma_{1 0}^{0 1} + \Gamma_{1 0}^{1 1}-\left(\Gamma_{0 1}^{0 0}+\Gamma_{0 1}^{1 0} \right),
\end{eqnarray}
With help of the relationships \cite{Schutz}:
\begin{eqnarray}
\label{eq.0.8}
&-&\langle n_{m}H_{m-1,m}\rangle= A_{1}+B_{1}\langle n_{m-1}\rangle-C_{1}\langle n_{m}\rangle+ D_{1}\langle n_{m-1} n_{m}\rangle \;\;,\;\;\nonumber\\
&-&\langle n_{m}H_{m,m+1}\rangle= A_{2}+B_{2}\langle n_{m+1}\rangle-C_{2}\langle n_{m}\rangle+ D_{2}\langle n_{m} n_{m+1}\rangle 
\end{eqnarray}
The equation of motion of the density at site $m$ (on a periodic chain) reads:
\begin{eqnarray}
\label{eq.0.10}
\frac{d}{dt}\langle n_{m} \rangle (t) = \frac{d}{dt}\langle \widetilde\chi|n_{m}e^{-Ht}|P(0)\rangle &=& A_{1}+A_{2}+B_{1}\langle n_{m-1} \rangle (t)+B_{2}\langle n_{m+1} \rangle (t)  \nonumber\\
&-&(C_{1}+C_{2})\langle n_{m} \rangle (t) +
D_{1}\langle n_{m-1}n_{m}\rangle (t) +D_{2}\langle n_{m} n_{m+1} \rangle(t)
\end{eqnarray}
In order to illustrate the physical meaning of the models studied in this work, let us consider the latter at the {\it mean-field} level (in the continuum limit), i.e. we assume first:
$\langle n_{x}(t)\rangle \rightarrow \rho_{MF}(x,t)$ and $\langle n_{x}n_{x\pm1}\rangle(t)\simeq (\rho_{MF}(x,t))^{2}$.
At this level of approximation, we note (see the section VIII) that the equation of motion (\ref{eq.0.10}) of some {\it microscopic} reaction-diffusion models studied in this work, provided that $D_{1}<0$ and $D_{2}<0$, is a non-linear
 partial-differential equation of {\it Fisher-type}\cite{Fisher,Murray,Riordan}:
\begin{eqnarray}
\label{eq.0.11}
\frac{\partial}{\partial t}\widetilde{\rho}_{MF}(x,t)= B\frac{\partial^{2}}{\partial x^{2}}\widetilde{\rho}_{MF}(x,t)+k_{1}\widetilde{\rho}_{MF}(x,t)-k_{2} (\widetilde{\rho}_{MF}(x,t))^{2},
\end{eqnarray}
with $\widetilde{\rho}_{MF}(x,t)\equiv\rho_{MF}(x,t)-\phi $.
When $A_{1}=A_{2}=A\;,B_{1}=B_{2}=B\;, C_{1}=C_{2}=C$, and $D_{1}=D_{2}=D<0$, we have  $2D\phi\equiv (C-B)+\sqrt{(B-C)^{2}-4AD}$,  $k_{1}\equiv 2\sqrt{(B-C)^{2}-4AD}>0$ which denotes the constant describing the {\it growth}  and $k_{2}\equiv-2D>0$ is the constant describing the {\it saturation} according to the local dynamics \cite{Fisher,Murray,Riordan}.

Fisher equation (\ref{eq.0.11}) admits two homogeneous steady-states, namely: $\widetilde{\rho}_{MF}(x,\infty)\equiv\widetilde{\rho}_{MF}(\infty) =k_{1}/k_{2}$, which is stable and another, unstable, steady-state: $\widetilde{\rho}_{MF}(\infty)=0$.

Although Fisher equation (\ref{eq.5.11}) cannot be solved exactly, it is known \cite{Murray,Riordan}  that the approach towards the steady-state from  inhomogeneous initial states (e.g. $\widetilde{\rho}_{MF}(x,0)=\frac{k_{1}}{k_{2}}\Theta(x_{0}-x)$, where $\Theta(x')=1$ if $x'>0$ and $\Theta(x')=0$ otherwise) is characterized by a {\it wave-front}: $\widetilde{\rho}_{MF}(x,t)=f(x-ct)$ propagating with a celerity $c\geq 2\sqrt{k_{1}B}$ and satisfying the non-linear differential equation: $B\frac{d^{2}}{dz^{2}}f(z)+c\frac{d}{dz}f(z)+k_{1}f(z)-k_{2}(f(z))^{2}=0$\cite{Murray}.

In this work we obtain {\it exact expression} for the density from the $N-$body description of some reaction-diffusion models (on finite and periodic lattice), for which $D_{1}=D_{2}\neq 0$ and for which, the mean-field description in the continuum limit is given by non-linear partial-differential equation of Fisher-type (\ref{eq.0.11}). Therefore, with help of the (microcopic) exact results obtained in section V and VI,  we are able, in section VIII,  to discuss the validity of the  mean-field description  {\it \`a la Fisher}. 

\section{The string function}
In this section we introduce the quantity which is the key to our  analysis, i.e. the string function $S_{x,y}(t)$. We also derive the constraints for which the equation of motion of the latter is a closed hierarchy. In the sequel, we solve the latter providing 
the density of particles and some two-point correlation functions.

Instead of considering  the {\it standard} ``empty-interval function'' \cite{Doering,benA,Hinrichsen,Peschel,Simon}, we focus here on the more general form ($1\leq x\leq y\leq x+L$):
\begin{eqnarray}
\label{eq.1.1}
S_{x,y}(t)\equiv\langle (a-bn_{x}) (a-bn_{x+1}) \dots (a-bn_{y-2}) (a-bn_{y-1}) \rangle(t),
\end{eqnarray}
where $a$ and $b$ are non-vanishing numbers.

This expression reduces to the {\it empty-interval function} when $a=b=1$ \cite{Doering,benA,Hinrichsen,Peschel,Simon}. Hereafter we will derive the equation of motion of the quantity $S_{x,y}(t)$ and determine which constraints are {\it necessary} and {\it sufficient} to close the latter. An alternative approach would consist in considering the  {\it empty-interval function} (with $a=b=1$ in (\ref{eq.1.1})) and obtaining a solution of some related reaction-diffusion model via a similarity transformation. This approach has been extensively studied for the free-fermion models \cite{Hinrichsen,Simon} where solutions of RD systems are obtained from the solution of the (free-fermion) diffusion-coagulation model. The latter approach is investigated in section VII. 
For $1\leq x\leq y<x+L$, the equation of motion of $S_{x,y}(t)$ reads: 
\begin{eqnarray}
\label{eq.1.2}
\frac{d}{dt}S_{x,y}(t)&=&
-\left\langle (a-bn_{x})H_{x-1,x}(a-bn_{x+1}) \dots (a-b n_{y-1}) 
\right\rangle(t) -\left\langle (a-bn_{x})(a -b n_{x+1}) \dots (a-b n_{y-1}) 
  H_{y-1,y}\right\rangle(t) \nonumber\\&-&\sum_{j=x}^{y-2}\left\langle (a-bn_{x})(a-b n_{x+1})
 \dots (a-bn_{y-1})   H_{j,j+1}\right\rangle(t)
\end{eqnarray}
If the following five constraints are fulfilled, the dynamics of
 $S_{x,y}(t)$ is described by a closed hierarchy of equations:
\begin{eqnarray}
\label{eq.1.3}
(1):\;&& aD_{1}=-bB_{1}; \;\; (2):\; aD_{2}=-bB_{2};\nonumber\\
(3-5):\;&& \Gamma_{0 0}^{0 0}+\left(\frac{a-b}{a}\right)(\Gamma_{0 0}^{1 0}+
\Gamma_{0 0}^{0 1} )+ \left(\frac{a-b}{a}\right)^{2} \Gamma_{0 0}^{1 1}\nonumber\\
&=&\frac{a}{b}(\Gamma_{0 0}^{0 0}-\Gamma_{1 0}^{0 0})-\left(\frac{a-b}{b}\right)
(\Gamma_{1 0}^{0 1}+\Gamma_{1 0}^{1 0}-\Gamma_{0 0}^{1 0}-\Gamma_{0 0}^{0 1})
-\frac{(a-b)^{2}}{ab}(\Gamma_{1 0}^{1 1}-\Gamma_{0 0}^{1 1})\nonumber\\
&=&\frac{a}{b}(\Gamma_{0 0}^{0 0}-\Gamma_{0 1}^{0 0})-\left(\frac{a-b}{b}\right)
(\Gamma_{0 1}^{0 1}+\Gamma_{0 1}^{1 0}-\Gamma_{0 0}^{1 0}-\Gamma_{0 0}^{0 1})
-\frac{(a-b)^{2}}{ab}(\Gamma_{0 1}^{1 1}-\Gamma_{0 0}^{1 1})\nonumber\\
&=&\left(\frac{a}{b}\right)^{2}(\Gamma_{1 1}^{0 0}+\Gamma_{0 0}^{0 0}-
\Gamma_{1 0}^{0 0} -\Gamma_{0 1}^{0 0} ) +\frac{a(a-b)}{b^{2}}(\Gamma_{1 1}^{1 0} +
\Gamma_{1 1}^{0 1}+\Gamma_{0 0}^{1 0}+ \Gamma_{0 0}^{0 1}-\Gamma_{0 1}^{1 0}-
 \Gamma_{1 0}^{0 1}-\Gamma_{1 0}^{1 0} -\Gamma_{0 1}^{0 1})\nonumber\\&+&
\left(\frac{a-b}{b}\right)^{2}(\Gamma_{1 1}^{1 1}+\Gamma_{1 1}^{0 0}-
\Gamma_{1 0}^{1 1}-\Gamma_{0 1}^{1 1})
\end{eqnarray}
The interesting point is that this five constraints are generally independent of the previous ones (\ref{eq.0.7}). Therefore, in general, models which are solvable via the approach presented here are not on the $10-$parameters manifold described by (\ref{eq.0.7}) where the equation of motion of  correlation functions are closed.

When the constraints (\ref{eq.1.3}) are fulfilled, the equation of motion of the {\it string-function} $S_{x,y}(t)$ on a periodic lattice of $L$ sites is the following (for $aC_{1}\neq bA_{1}$ and $aC_{2}\neq bA_{2}$)
\footnote{If $aC_{1}= bA_{1}$ and $aC_{2}= bA_{2}$, 
$\frac{b}{a}=\frac{C_{1}}{A_{1}}=\frac{C_{2}}{A_{2}}$
and the equation of motion of $S_{x,y}$ reads:
$\left\{
\begin{array}{l l}
\frac{d}{dt}S_{x,y}(t)=
-\frac{D_{1}}{b}S_{x-1,y}(t)-\frac{D_{2}}{b}S_{x,y+1}(t)
-(\gamma+\delta)S_{x,y}(t)-(y-x-1)\delta S_{x,y},\;\;
 (1\leq x<y<x+ L)\\
\frac{d}{dt}S_{x,x+L}(t)=-L\delta  S_{x,x+L}(t)
\end{array}
\right. $
\\
,where $\gamma+\delta\equiv B_{1}+B_{2}+C_{1}+C_{2}$, and
$\delta\equiv-\left[\Gamma_{0 0}^{0 0}+\left(\frac{a-b}{a}\right)(\Gamma_{0 0}^{1 0}+
\Gamma_{0 0}^{0 1} )+ \left(\frac{a-b}{a}\right)^{2} \Gamma_{0 0}^{1 1} \right]$. An example of model which dynamics is described by such a system of equations (with $a=b$) is the {\it random sequential adsorbtion} (RSA) process of dimers: $\emptyset\emptyset\longrightarrow {\cal A}{\cal A} \cite{Evans}$.
 }:
\begin{eqnarray}
\label{eq.1.4}
\left\{
\begin{array}{l l l}
\frac{d}{dt}S_{x,y}(t)=(aC_{1}-bA_{1})S_{x+1,y}(t)+(aC_{2}-bA_{2})S_{x,y-1}(t)
-\frac{D_{1}}{b}S_{x-1,y}(t)-\frac{D_{2}}{b}S_{x,y+1}(t)\\-(B_{1}+B_{2}+C_{1}+C_{2})S_{x,y}(t)
+\left[\Gamma_{0 0}^{0 0}+\left(\frac{a-b}{a}\right)(\Gamma_{0 0}^{1 0}+
\Gamma_{0 0}^{0 1} )+ \left(\frac{a-b}{a}\right)^{2} \Gamma_{0 0}^{1 1} \right](y-x-1)S_{x,y}(t);\;\; (1\leq x<y<x+L)\\
\frac{d}{dt}S_{x,x+L}(t)=L\left[\Gamma_{0 0}^{0 0}+\left(\frac{a-b}{a}\right)(\Gamma_{0 0}^{1 0}+
\Gamma_{0 0}^{0 1} )+ \left(\frac{a-b}{a}\right)^{2} \Gamma_{0 0}^{1 1} \right]  S_{x,x+L}(t)\\
S_{x,x}(t)=1,
\end{array}
\right.
\end{eqnarray}
where the {\it boundary condition} $S_{x,x}(t)=1$ is obtained from the requirement that $\frac{d}{dt}S_{x,x+1}(t)=-b \frac{d}{dt}\langle n_{x}(t)\rangle$.

Let us note that when $a=b=1$, we recover the usual constraints of the IPDF method:
$\Gamma_{1 0}^{0 1}=\Gamma_{1 1}^{0 1}$, $\Gamma_{0 1}^{1 0}=\Gamma_{1 1}^{1 0}$, 
$\Gamma_{1 0}^{0 0}=\Gamma_{0 1}^{0 0}=\Gamma_{1 1}^{0 0}=0$.
\section{Solution of the equation of motion of the string-function }
Equations of motion for the string-function have been intensively studied for free-fermion systems both in the continuum limit \cite{Doering,benA} and on discrete lattices (periodic and open boundary conditions) \cite{Hinrichsen,Peschel,Simon}. However, for systems which cannot be mapped onto free-fermion systems, a few results have been obtained: Doering and ben-Avraham \cite{Doering} have obtained the stationary concentration and the relaxation spectrum of a reversible diffusion-coagulation model on an infinite lattice in the continuum limit. Later, Peschel et {\it al.} \cite{Peschel} have studied the relaxation spectrum of the Fourier transform of the string-function on a periodic lattice with help of the {\it conventional} IPDF method.

It is useful to introduce the following notations:
\begin{eqnarray}
\label{eq.2.1}
\frac{\alpha_{1}}{2}\equiv aC_{1}-bA_{1}\neq 0;\; \frac{\alpha_{2}}{2}\equiv aC_{2}-bA_{2}\neq 0;\;
\frac{\beta_{1}}{2}\equiv -\frac{D_{1}}{b}=\frac{B_{1}}{b};\; \frac{\beta_{2}}{2}\equiv -\frac{D_{2}}{b}=\frac{B_{2}}{b}\nonumber\\
\delta\equiv-\left[\Gamma_{0 0}^{0 0}+\left(\frac{a-b}{a}\right)(\Gamma_{0 0}^{1 0}+
\Gamma_{0 0}^{0 1} )+ \left(\frac{a-b}{a}\right)^{2} \Gamma_{0 0}^{1 1} \right];\;
\gamma+\delta\equiv B_{1}+B_{2}+C_{1}+C_{2}
\end{eqnarray}
Herefater we solve the equation of motion (\ref{eq.1.4}) for the case  $\delta \neq 0$ ($\delta=0$ corresponds to the {\it free-fermion case}), $D_{1}\neq 0$ and $D_{2}\neq 0$\footnote{
The case where $D_{1}=D_{2}=0$
 corresponds to the situation where the equations of motion of {\it all} the correlation functions close (see,(\ref{eq.0.7}), (\ref{eq.0.10})) .
For a translationally-invariant system with initial density $\rho(0)$and with $B_{1}+B_{2}\neq C_{1}+C_{2}$, the density simply reads:
$\langle n_{j}(t)\rangle=\frac{A_{1}+A_{2}}{C_{1}+C_{2}-(B_{1}+B_{2})}+
\left(\rho(0)-\frac{A_{1}+A_{2}}{C_{1}+C_{2}-(B_{1}+B_{2})}\right)e^{-\left(C_{1}+C_{2}-(B_{1}+B_{2})\right)t}.$}, with the additional condition, $\alpha_{2}=\alpha_{1}\equiv \alpha\neq 0$ and $\beta_{1}=\beta_{2}\equiv\beta\neq 0$, which corresponds to the restriction to {\it unbiased} systems.

We also introduce the following auxiliary function: ${\cal R}_{x,y}(t)\equiv \mu^{x-y}S_{x,y}(t)$, where we choose $\mu\equiv\sqrt{\frac{\alpha_{1}}{\beta_{2}}}=
\sqrt{\frac{\alpha}{\beta}}$ and $q=\sqrt{\alpha_{1}\beta_{2}}=\sqrt{\alpha \beta}$ and, we solve (\ref{eq.1.4}) for the general case where $q\neq 0$ (Note that  $\beta_{i}=0 \leftrightarrow D_{i}=B_{i}=q=0$, with $i\in (1,2)$; see footnote $(2)$ ). With these notations, the equation of motion (\ref{eq.1.4}) becomes:
\begin{eqnarray}
\label{eq.2.2}
\left\{
\begin{array}{l l l}
\frac{d}{dt}{\cal R}_{x,y}(t)=\frac{q}{2}\sum_{e=\pm1}\left\{ {\cal R}_{x+e,y}(t)+ {\cal R}_{x,y+e}(t) \right\}-\gamma {\cal R}_{x,y}(t)-(y-x)\delta {\cal R}_{x,y}(t);\;\; (1\leq x<y<x+L)\\
\frac{d}{dt}{\cal R}_{x,x+L}(t)=- L\delta {\cal R}_{x,x+L}(t)\\
{\cal R}_{x,x}(t)=1,
\end{array}
\right.
\end{eqnarray}
The stationary solution of the system (\ref{eq.2.2}) is obtained with help of the properties of Bessel functions of first and second kind, respectively, 
$J_{\nu}(z)$ and $Y_{\nu}(z)$ \cite{tables}. In fact, the structure of the equation (\ref{eq.2.2}) for $1\leq x<y<L$ suggests the Ansatz: $R_{x,y}(\infty)=\widetilde{A}_{L}J_{y-x+\omega}(2q/\delta)
+\widetilde{B}_{L}Y_{y-x+\omega}(2q/\delta) $. Inserting this expression into (\ref{eq.2.2}), we obtain $\omega=\gamma/\delta$. Therefore we have:
\begin{eqnarray}
\label{eq.2.3}
R_{x,y}(\infty)=\widetilde{A}_{L}J_{y-x+\gamma/\delta}(2q/\delta)
+\widetilde{B}_{L}Y_{y-x+\gamma/\delta}(2q/\delta)
\end{eqnarray}
The quantities $\widetilde{A}_{L}$ and $\widetilde{B}_{L}$ are determined with help of the boundary conditions: $R_{x,x}(\infty)=1=\widetilde{A}_{L}J_{1+\gamma/\delta}(2q/\delta)
+\widetilde{B}_{L}Y_{1+\gamma/\delta}(2q/\delta) $ and
 $R_{x,x+L}(\infty)=\widetilde{A}_{L}J_{L+\gamma/\delta}(2q/\delta)
+\widetilde{B}_{L}Y_{L+\gamma/\delta}(2q/\delta)=0$.
It follows
\begin{eqnarray}
\label{eq.2.4}
\widetilde{A}_{L}=-\frac{Y_{L+\frac{\gamma}{\delta}}(\frac{2q}{\delta})}
{J_{L+\frac{\gamma}{\delta}}(\frac{2q}{\delta})Y_{\frac{\gamma}{\delta}}(\frac{2q}{\delta})-
Y_{L+\frac{\gamma}{\delta}}(\frac{2q}{\delta})J_{\frac{\gamma}{\delta}}(\frac{2q}{\delta})}\;,\;
\widetilde{B}_{L}= \frac{J_{L+\frac{\gamma}{\delta}}(\frac{2q}{\delta})}
{J_{L+\frac{\gamma}{\delta}}(\frac{2q}{\delta})Y_{\frac{\gamma}{\delta}}(\frac{2q}{\delta})-
Y_{L+\frac{\gamma}{\delta}}(\frac{2q}{\delta})J_{\frac{\gamma}{\delta}}(\frac{2q}{\delta})},
\end{eqnarray}
 which, with (\ref{eq.2.3}), provides the stationary expression of 
${\cal R}_{x,y}(\infty)$ and thus we obtain the stationary expression of the 
string-function:
\begin{eqnarray}
\label{eq.2.5}
S_{x,y}(\infty)=\mu^{y-x}\left[\frac{J_{L+\gamma/\delta}(2q/\delta)Y_{y-x+\gamma/\delta}(2q/\delta)- Y_{L+\gamma/\delta}(2q/\delta)J_{y-x+\gamma/\delta}(2q/\delta) }
{J_{L+\gamma/\delta}(2q/\delta)Y_{\gamma/\delta}(2q/\delta)- Y_{L+\gamma/\delta}(2q/\delta)J_{\gamma/\delta}(2q/\delta)}\right],
\end{eqnarray}

To solve the dynamical part of (\ref{eq.2.2}), we seek a solution of the form: ${\cal R}_{x,y}(t)-{\cal R}_{x,y}(\infty)=\sum_{\lambda}r_{y,x}^{\lambda}e^{-\lambda q t} $, which leads to the following difference equation:
\begin{eqnarray}
\label{eq.2.6}
r_{y,x+1}^{\lambda}+r_{y-1,x}^{\lambda}+r_{y,x-1}^{\lambda}+r_{y+1,x}^{\lambda}+2\left(\lambda -\{\frac{\gamma+(y-x)\delta}{q}\}\right)r_{y,x}^{\lambda}=0,
\end{eqnarray}
with the boundary conditions:
\begin{eqnarray}
\label{eq.2.7}
(L\delta-\lambda q)r_{x,x+L}^{\lambda}=0\;\; \text{ and}\;\;
r_{x,x}^{\lambda}=0,
\end{eqnarray}
Introducing the notation: $E\equiv\frac{q\lambda -\gamma}{\delta}$, (\ref{eq.2.6}) admits as solution: $r_{x,y}^{\lambda}=\widetilde{{\cal A}}J_{y-x-E}(2q/\delta)+\widetilde{{\cal B}}Y_{y-x-E}(2q/\delta)$, where $\widetilde{{\cal A}}, \widetilde{{\cal B}}$ and the (relaxation-)spectrum $\{E\}$ are determined from the boundary conditions (\ref{eq.2.7}), which imply:
\begin{eqnarray}
\label{eq.2.8}
\left\{
\begin{array}{l l}
\widetilde{{\cal A}}J_{-E}(2q/\delta)+\widetilde{{\cal B}}Y_{-E}(2q/\delta)=0\\
\widetilde{{\cal A}}J_{L-E}(2q/\delta)+\widetilde{{\cal B}}Y_{L-E}(2q/\delta)=0,
\end{array}
\right.
\end{eqnarray}
The only non-trivial solution of this system (for which $\widetilde{{\cal A}}\neq 0$ and $\widetilde{{\cal B}}\neq 0$) requires
\begin{eqnarray}
\label{eq.2.9}
J_{L-E}(2q/\delta)Y_{-E}(2q/\delta)-J_{-E}(2q/\delta)Y_{L-E}(2q/\delta)=0, 
\end{eqnarray}
or equivalently in terms of {\it Lommel function} \cite{tables}:
\begin{eqnarray}
\label{eq.2.10}
R_{L-1,1-E}(2q/\delta)=0, 
\end{eqnarray}
which admits $L-1$ zeroes \cite{Peschel,Stey,tables} with degeneracy $L$. 
The latter are symmetrically distributed around $\frac{L}{2}$ (which is an {\it eigenvalue} if $L$ is even).
To obtain the complete set of $L(L-1)+1$ eigenvalues, i.e., the complete relaxation-spectrum $\{E_{i}\}, i=1,\dots,L$ of the string-function (which {\it has not to be confused with the spectrum of the stochastic Hamiltonian $H$}), one has to take into account the {\it eigenvalue} $q\lambda=L\delta$ directly obtained from the boundary condition (\ref{eq.2.7}). Notice that in $\{E_{i}\}$ the index $i=1,\dots,L$ labels the $L$ {\it distinct eigenvalues} forming the relaxation-spectrum. 
In order to have some more insight into the {\it relaxation-spectrum} $\{E_{i}\}, i=1,\dots,L$ of the string-function $S_{x,y}(t)$ , we can take advantage from the fact that the following eigenvalue-problem:
\begin{eqnarray}
\label{eq.2.11}
\left\{
\begin{array}{l l }
(E-n)F_{n}=V\left(F_{n-1}+F_{n+1}\right)\;;(1\leq n<L)\\
F_{0}=F_{L}=0,
\end{array}
\right.
\end{eqnarray}
admits as eigenvalues the ($L-1$) zeroes of the {\it Lommel function}: $R_{L-1,1-E}(2V)=0$ \cite{Stey,Peschel}.
Therefore, choosing $V=\frac{q}{\delta}$, the problem of determining the relaxation spectrum is reformulated as that of solving the eigenvalue-problem (\ref{eq.2.11})
${\cal M}|{\cal F}\rangle\rangle=E|{\cal F}\rangle\rangle$, where ${\cal M}$ is a $(L-1)\times (L-1)$ symmetric (but not {\it hermitian} when $q$ has an imaginary part) tridiagonal matrix and $|{\cal F}\rangle\rangle$ is a $(L-1)$-components  column-vector: $|{\cal F}\rangle\rangle\equiv (F_{n=1}\; F_{2}\; \dots  F_{L-1})^{T}$. The general form of the matrix ${\cal M}$ is the following:
\begin{eqnarray}
\label{eq.2.12}
{\cal M} =
\left(
 \begin{array}{c c c c c c c}
 1 &  q/\delta &   0  & \dots &\dots & \dots &   0  \\
 q/\delta  &  2 &  q/\delta &  0 &\dots & \dots &  0  \\
 0 &  q/\delta  &  3 &  q/\delta & 0 &\dots  &  0  \\
 0 & \ddots &\ddots & \ddots & \ddots  & \ddots & \vdots\\
\vdots &\ddots & \ddots & \ddots & \ddots & \ddots &  0\\
 0 & \dots &  0 & \ddots &   q/\delta&  (L-2) &   q/\delta\\
 0 & \dots &\dots & \dots & 0 &  q/\delta&  (L-1) 
 \end{array}
\right)\
\end{eqnarray}
For {\it small} systems the $(L-1)$ distinct eigenvalues $\{E_{i}\}$ of ${\cal M}$, can be computed analytically. For $L=6$, with $V=\frac{q}{\delta}$, we have $\{E_{i}\}=\left\{3,3\pm\sqrt{\frac{5+4V^{2}\pm\sqrt{9+24V^{2}+4V^{4}}}{2}}\right\}$, where we still have to take into account the {\it additional eigenvalue} $q\lambda=L\delta$. The spectrum depends on the size $L$ of the system and this is in particular the case for $E^{\ast}\equiv min_{E}\{E\}\equiv \epsilon_{L}$, the {\it smallest} eigenvalue, which governs the long-time behaviour of the system.
 For larger matrices we had to proceed numerically and for $L\gg 1$, $\epsilon_{L}\rightarrow \epsilon_{\infty}$, and  $E^{\ast}$ is a constant: $E^{\ast}=\epsilon_{\infty}$. For $L=6$, we have the exact result $\epsilon_{L=6}=3-\sqrt{\frac{5+4V^{2}+\sqrt{9+24V^{2}+4V^{4}}}{2}}$. This expression can be considered as an excellent approximation for systems of size  $L\gg 1$ and in particular for $\epsilon_{\infty}$.
 
 Therefore, the long-time dynamics (of large systems, with $L\gg 1$) is governed by the eigenvalue ($V\equiv q/\delta$):
\begin{eqnarray}
\label{eq.2.13.0}
E^{\ast}=\epsilon_{L}\simeq3-\sqrt{\frac{5+4(\frac{q}{\delta})^{2}+
\sqrt{9+24(\frac{q}{\delta})^{2}+4(\frac{q}{\delta})^{4}}}{2}}=\epsilon_{L=6}
\end{eqnarray}
, i.e.
\begin{eqnarray}
\label{eq.2.13}
q\lambda^{\ast}=E^{\ast}\delta+\gamma =\epsilon_{L}\delta+\gamma
\end{eqnarray}
This expression provides the {\it inverse of the relaxation-time} of the model under consideration.

Having obtained the relaxation spectrum and the expression of $r_{y,x}^{\lambda}$, the complete expression for the string-function follows as:
\begin{eqnarray}
\label{eq.2.14}
S_{x,y}(t)-S_{x,y}(\infty)=\mu^{y-x}\sum_{E_{i}}{\cal A}_{E_{i}}e^{-(E_{i}\delta+\gamma)t}\left[J_{y-x-E_{i}}(2q/\delta)Y_{L-E_{i}}(2q/\delta) -
Y_{y-x-E_{i}}(2q/\delta)J_{L-E_{i}}(2q/\delta)
\right],
\end{eqnarray}

Here for simplicity we consider the translationally-invariant (but {\it not} necessarily uncorrelated) situation, when $S_{x,y}(t)=S_{y-x}(t)$. In this case,  the coefficients ${\cal A}_{E_{i}}$ are obtained from the initial condition according to :
\begin{eqnarray}
\label{eq.2.15.0}
{\cal A}_{E_{i}}=\sum_{j,n=1}^{L}\left[{\cal N}^{-1}\right]_{i,j}
\left(J_{n-E_{j}}(2q/\delta)Y_{L-E_{j}}(2q/\delta) -Y_{n-E_{j}}(2q/\delta)J_{L-E_{j}}(2q/\delta) \right)^{\ast}\left(S_{n}(0)-S_{n}(\infty)\right)\mu^{-n},
\end{eqnarray}
where ${\cal N}$ is an hermitian $L\times L$ matrix (see (\ref{eq.2.20.0})).

To clarify this point let us introduce the following {\it vectors} of the Hilbert space ${\bf C}^{L}$ (with the {\it usual} scalar product):
\begin{eqnarray}
\label{eq.2.16.0}
 |S\rangle\rangle\equiv\left((S_{1}(0)-S_{1}(\infty))\mu^{-1} \dots (S_{L}(0)-S_{L}(\infty))\mu^{-L})\right)^{T}
\end{eqnarray}
 and
\begin{eqnarray}
\label{eq.2.17.0}
|V_{E_{j}}\rangle\rangle\equiv
\left(
 \begin{array}{c c c c}
 \left(J_{1-E_{j}}(2q/\delta)Y_{L-E_{j}}(2q/\delta) -Y_{1-E_{j}}(2q/\delta)J_{L-E_{j}}(2q/\delta) \right) \\
\vdots\\
\left(J_{L-1-E_{j}}(2q/\delta)Y_{L-E_{j}}(2q/\delta) -Y_{L-1-E_{j}}(2q/\delta)J_{L-E_{j}}(2q/\delta) \right)  \\
0 \\
 \end{array}\right)\
\end{eqnarray}
In a vectorial formulation, the  coefficients ${\cal A}_{E_{j}}$ are
 obtained from the initial state of the system according to:
\begin{eqnarray}
\label{eq.2.18.0}
|S\rangle\rangle=\sum_{j=1}^{L}{\cal A}_{E_{j}}|V_{E_{j}}\rangle\rangle
\end{eqnarray}
Solving this equation, we formally obtain the expression for the coefficients
${\cal A}_{E_{j}}$ :
\begin{eqnarray}
\label{eq.2.19.0}
{\cal A}_{E_{j}}=\sum_{j=1}^{L}\left[{\cal N}^{-1}\right]_{i,j}
\langle\langle V_{E_{j}}|S\rangle\rangle,
\end{eqnarray}
where ${\cal N}$ is an hermitian $L\times L$ matrix which entries read:
\begin{eqnarray}
\label{eq.2.20.0}
\left[{\cal N}\right]_{i,j}\equiv\langle\langle V_{E_{i}}| V_{E_{j}}\rangle\rangle   =\sum_{n=1}^{L}&&\left(J_{n-E_{i}}(2q/\delta)Y_{L-E_{i}}(2q/\delta) -Y_{n-E_{i}}(2q/\delta)J_{L-E_{i}}(2q/\delta) \right)^{\ast}\nonumber\\ &\times& \left(J_{n-E_{j}}(2q/\delta)Y_{L-E_{j}}(2q/\delta) -Y_{n-E_{j}}(2q/\delta)J_{L-E_{j}}(2q/\delta) \right)
\end{eqnarray}
With help of the expression of the string-function (\ref{eq.2.14}), we can compute the exact density of particles at site $x$:

\begin{eqnarray}
\label{eq.2.15}
\langle n_{x}(t)\rangle=\frac{a-S_{x,x+1}(t)}{b}
\end{eqnarray}

In the non-translationally-invariant situation, we would proceed in a similar manner, but we would have to work with vectors of the Hilbert space ${\bf C}^{L(L-1)+1}$. In this case we would have to take into account the degeneracy of the eigenvalues of ${\cal M}$ in order to compute the $L(L-1)+1$ components ${\cal A}_{E}$ appearing in (\ref{eq.2.14}). This is achieved by replacing $E_{i}$ with $E_{i,d}$ in (\ref{eq.2.14}), where the index $d$ labels the degeneracy of the eigenvalues $E_{i}, i=1,\dots,L$.

From  (\ref{eq.2.15}), we can also obtain the expression of the non-instantaneous two-point correlation functions $\langle n_{x}(t)n_{x_{0}}(0)\rangle$. We should take into account the initial state (\ref{eq.2.15}) $|P'(0)\rangle\equiv n_{x_{0}}|P(0)\rangle$ instead of $|P(0)\rangle$.

With (\ref{eq.2.14}), we can also compute the {\it instantaneous} nearest-neighbour (two-point) correlation functions:
\begin{eqnarray}
\label{eq.2.16}
\langle n_{x}n_{x+1}\rangle(t)=\frac{a^{2}+S_{x,x+2}(t)-a(S_{x,x+1}(t)+S_{x+1,x+2}(t))}{b^{2}},
\end{eqnarray}
Although the present approach could be formally extended to obtain the exact and closed equation of motion of other {\it string-like functions}, which (eventual) resolution would provide the exact expressions of all instantaneous two-point correlation functions; in practice, such equations turn out to be extremely difficult to solve. The only cases where the whole hierarchy of equation has been completely solved are the free-fermion models. On the discrete (and finite) lattice these solutions were obtained by Hinrichsen {\it et al.} \cite{Hinrichsen} and in the continuum limit (for the diffusion-coagulation ${\cal A}{\cal A}\longrightarrow {\cal A}$ model) by ben-Avraham \cite{benA}. Unfortunately it is known that the latter approaches cannot be extended to systems which cannot be mapped onto free-fermion systems \cite{Hinrichsen,benA}.   Here we prefer to take advantage of the quantities $\langle n_{x}(t)\rangle$ and $\langle n_{x}n_{x+1}(t)\rangle$ and $S_{x,y}(t)$,
 which we can compute exactly to obtain approximative instantaneous two-point correlation functions of the systems obeying the constraints
 (\ref{eq.1.3}), which  associated string-function $S_{x,y}(t)$ obeys the equation of motion (\ref{eq.1.4}).

For technical convenience, we consider the {\it translationally-invariant} situation (thus $\langle n_{x}(t)\rangle=\rho(t)$) and expanding  the string function we have:

\begin{eqnarray}
\label{eq.2.17.2}
&&\left(\frac{a}{b}\right)^{2}\left[a^{-(y-x)}S_{y-x}(t)+\frac{b(y-x)}{a}\rho(t)\right]=\sum_{j=1}^{y-1}(y-x-j)\langle n_{j_{1}} n_{j_{1}+j}\rangle(t)
+\dots\nonumber\\&+&\left(\frac{b}{a}\right)^{m-2}
\sum_{y-x>j_{1}>j_{2}>\dots>j_{m}}\langle n_{j_{1}}n_{j_{2}} \dots n_{j_{m}}\rangle(t)+\dots+b^{y-x-2}
\langle n_{x}n_{x+1}\dots n_{y-1}
\rangle(t)
\end{eqnarray}
From equation (\ref{eq.2.17.2}), it is possible to obtain
{\it exact expressions} relating a $two-point$ correlation function with known quantities and {\it higher order correlation functions}.
As an illustration, let us first consider the case where  $y-x=3$. Equation (\ref{eq.2.17.2}) implies:
\begin{eqnarray}
\label{eq.2.17.3}
\langle n_{x}n_{x+2}\rangle(t)-\frac{b}{a}\langle n_{x}n_{x+1}n_{x+2}\rangle(t)=
\frac{S_{x,x+3}}{ab^{2}}-\frac{a^{2}}{b}+\frac{3a\rho(t)}{b}-2\langle n_{x}n_{x+1}\rangle(t)
\end{eqnarray}
For  $y-x=4$, (\ref{eq.2.17.2}) with help of (\ref{eq.2.17.3}) leads to:
\begin{eqnarray}
\label{eq.2.17.4}
\langle n_{x}n_{x+3}\rangle(t)-\left(\frac{b}{a}\right)^{2}\langle n_{x}n_{x+1}n_{x+2}n_{x+3} \rangle(t)=
\frac{S_{x,x+4}(t)-2aS_{x,x+3}(t) }{(ab)^{2}}
+\left(\frac{a}{b}\right)^{2}-\frac{2a\rho(t)}{b}+ \langle n_{x}n_{x+1}\rangle(t)
\end{eqnarray}
This procedure can naturally be continued for every two-point correlation functions.
Therefore, for two-point correlation functions $\langle n_{x}n_{x+r}\rangle (t)$ of sites separated by a distance $r$, using recursively the relations previously derived for  $\langle n_{x}n_{x+r-1}\rangle (t)$, $\langle n_{x}n_{x+r-2}\rangle (t), \dots$  we obtain an equality relating  $\langle n_{x}n_{x+r}\rangle (t)$ and an unknown  $r+1$-point correlation function ($\langle n_{x}n_{x+1}\dots n_{x+r-1} n_{x+r}\rangle (t)$) to a combination of known quantities, as in (\ref{eq.2.17.3}) and (\ref{eq.2.17.4}). 

It is therefore possible to obtain {\it approximative} expressions for the correlation function within the {\it truncation approximation} (for $r$ even):
 $\langle n_{x}n_{x+1}\dots n_{x+r-1} n_{x+r}\rangle (t)\simeq
\langle n_{x}n_{x+1}\rangle (t)\dots\langle  n_{x+r-1} n_{x+r}\rangle (t)=
[\langle n_{x}n_{x+1}\rangle (t)]^{r/2}
$
and $\langle n_{x}n_{x+1}\dots n_{x+r-1} n_{x+r}\rangle (t)\simeq
[\langle n_{x}n_{x+1}\rangle (t)]^{(r-1)/2}\rho(t)$, for $r$ odd.

Within this {\it mean-field-like} approach,  we obtain the following approximative expressions for the two-point correlation functions:
\begin{eqnarray}
\label{eq.2.17.5}
\langle n_{x}n_{x+2}\rangle(t)&\simeq& \frac{S_{x,x+3}(t)}{ab^{2}}-\left(2+\frac{b\rho(t)}{a}\right)\langle n_{x}n_{x+1}\rangle(t)-\frac{3a\rho(t)}{b} -\left(\frac{a}{b}\right)^{2}\nonumber\\
\langle n_{x}n_{x+3}\rangle(t)&\simeq&
\frac{S_{x,x+4}(t)-2aS_{x,x+3}(t)}{(ab)^{2}}+\left(1-\left(\frac{b}{a}\right)^{2}\langle n_{x}n_{x+1}\rangle(t)\right)\langle n_{x}n_{x+1}\rangle(t)-\frac{2a\rho(t)}{b}+\left(\frac{a}{b}\right)^{2}
 \nonumber\\
\vdots
\end{eqnarray}
To conclude this section, let us comment on this recursive procedure.

 First of all, the {\it recursive} character of the method appears through the repeated use of  (\ref{eq.2.17.2}) and of the relations obtained for the other  two-point correlation functions. The advantages of this recursive mean-field-like method with respect to the traditional mean-field are the following:

(i) The procedure of {\it truncation} appears at the level of the three-point correlation functions.

(ii) This approach is based on the explicit knowledge of the quantities $S_{y-x}(t)$, $\rho(t)$ and $\langle n_{x}n_{x+1}\rangle(t)$.

(iii) This method does not give rise to non-linear partial differential equations and/or to non-linear self-consistent equations which are generally difficult to solve and which appear from traditional mean-field methods. Conversely, the approach presented here gives directly (after the {\it truncation} procedure) access to  the (approximative) expressions of the two-point correlation functions. 

\vspace{0.3cm}

It follows from the exact expression (\ref{eq.2.14}) of $S_{x,y}(t)$ that, for the models under consideration in this work, the related {\it string-function} approaches its steady-state exponentially fast, with an inverse relaxation-time given  by (\ref{eq.2.13}). This result is valid for an arbitrary initial state: the effect of the initial condition only appears through multiplicative factors ${\cal A}_{E}$. In the translationally-invariant situation we have the coefficients (\ref{eq.2.15.0}); other initial conditions do not  affect the exponential nature of the relaxation (\ref{eq.2.14}) with the inverse of relaxation-time (\ref{eq.2.13}).

\section{A model which cannot be solved directly from the {\it conventional} IPDF method}
In this section we consider a model which cannot be solved directly by the {\it conventional } IPDF method. A brief account of the study of this model has recently been reported  in \cite{Mobar}. Here we complete and develop this preliminary work.

The dynamics of the  model under consideration takes place on a finite and periodic lattice. When a particle and a vacancy are adjacent to each other, a {\it branching reaction} can take place and the particle ${\cal A}$ can give birth to an offspring (${\cal A}\emptyset\rightarrow {\cal A}{\cal A}$ and $\emptyset {\cal A}\rightarrow {\cal A}{\cal A}$) with rate $\Gamma_{1 0}^{1 1}=\Gamma_{0 1}^{1 1}$; 
another possible reaction is the {\it death} of the particle 
(${\cal A}\emptyset \rightarrow \emptyset\emptyset$ and $\emptyset {\cal A} \rightarrow \emptyset\emptyset$) with rate $\Gamma_{1 0}^{0 0}=\Gamma_{0 1}^{0 0}$. When two particles are adjacent, they can {\it coagulate} (${\cal A}{\cal A}\rightarrow {\cal A}\emptyset$ and ${\cal A}{\cal A}\rightarrow \emptyset {\cal A}$) with rate  $\Gamma_{1 1}^{1 0}= \Gamma_{1 1}^{0 1}$. In addition, when two vacancies are adjacent, a particle can appear ({\it birth} process, $\emptyset \emptyset\rightarrow {\cal A}\emptyset$ and $\emptyset \emptyset\rightarrow \emptyset {\cal A}$) with rate $\Gamma_{0 0}^{1 0}= \Gamma_{0 0}^{0 1}$.
The dynamics of this {\it branching-coagulation with birth and death processes} (BCBD) model can be encoded by the following reactions:
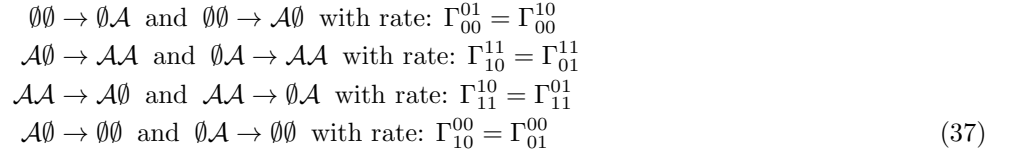
\begin{eqnarray}
\label{eq.3.1}
\emptyset \emptyset &\rightarrow& \emptyset {\cal A} \;\;\text{and} 
 \;\;\emptyset \emptyset \rightarrow  {\cal A} \emptyset \;\; \text{with rate: $\Gamma_{0 0}^{0 1}=\Gamma_{0 0}^{1 0}$} \nonumber\\
{\cal A} \emptyset &\rightarrow& {\cal A} {\cal A} \;\;\text{and} 
 \;\;\emptyset {\cal A} \rightarrow  {\cal A} {\cal A} \;\; \text{with rate: $\Gamma_{1 0}^{1 1}=\Gamma_{0 1}^{1 1}$} \nonumber\\
{\cal A} {\cal A} &\rightarrow& {\cal A} \emptyset \;\;\text{and} 
 \;\; {\cal A} {\cal A} \rightarrow  \emptyset {\cal A} \;\; \text{with rate: $\Gamma^{1 0}_{1 1}=\Gamma^{0 1}_{1 1}$} \nonumber\\
{\cal A} \emptyset &\rightarrow& \emptyset \emptyset \;\;\text{and} 
 \;\;  \emptyset {\cal A} \rightarrow  \emptyset \emptyset \;\; \text{with rate:
 $\Gamma_{1 0}^{0 0}=\Gamma_{0 1}^{0 0}$}
\end{eqnarray}
It should be emphasized that in this model, the effective motion of the particles is realized by successive processes of branching, coagulation, birth and death on neighbouring pairs of lattice sites, without {\it explicit} diffusion process. 

 The system described above can be viewed as an {\it epidemic model} where 
 particles can spontaneously appear/disappear,  have an offspring and coagulate. It can also be viewed as a generalization of the {\it voter} model \cite{Schutzrev}, where the presence/absence of particle is associated to an opinion (yes/no) and each site is associated to an human being. According to the dynamics of the model, each individual changes his opinion at a rate proportional to the opinion of his neighbours.

For the model under consideration $A_{1}=A_{2}=A=\Gamma_{0 0}^{1 0}$,
$B_{1}=B_{2}=B=\Gamma_{1 0}^{1 1}-\Gamma_{0 0}^{1 0}$, $C_{1}=C_{2}=C=\Gamma_{1 0}^{0 0}+\Gamma_{0 0}^{1 0}$, $D_{1}=D_{2}=D=\Gamma^{0 0}_{1 0}+\Gamma_{0 0}^{1 0}-(\Gamma_{1 0}^{1 1}+\Gamma_{1 1}^{1 0})$. If $D=0$, for the translationally-invariant situation with an initial density $\rho(0)$ of particles, we have (when $B\neq C$, see footnote $(2)$) $\langle n_{x}(t)\rangle=\frac{A}{C-B}+\left(\rho(0)-\frac{A}{C-B} \right) e^{-2(C-B)t}$. 

In this section we solve, with some restrictions on the reaction-rates, the above-mentioned model when  $D\neq 0$, i.e. in the case where  the equation of motion of the correlation functions of the model give rise to an open hierarchy (\ref{eq.0.10}). It has to be stressed that this model can be casted into a free-fermion form only  when $\Gamma_{1 0}^{1 1}=\Gamma_{1 0}^{0 0}$ and $\Gamma_{1 1}^{1 0}=\Gamma_{0 0}^{1 0}$(see \cite{Schutzrev} for a complete classification of {\it free-fermion} systems). Furthermore, this model cannot be solved (directly) by the {\it traditional} IPDF method (not applicable  \cite{Doering,benA,Hinrichsen,Peschel} in the presence of the processes ${\cal A} \emptyset \rightarrow \emptyset \emptyset\;;\; \emptyset {\cal A} \rightarrow \emptyset \emptyset $ and in the absence of processes  ${\cal A} \emptyset\rightarrow \emptyset {\cal A}\;;\; \emptyset {\cal A} \rightarrow {\cal A} \emptyset $; the latter should occur with the same rate as the coagulation rates \cite{Doering,benA,Hinrichsen,Peschel}). 
The idea  is to {\it choose} suitable $a$ and $b$ to close the equation of evolution of  $S_{x,y}(t)$. This is achieved by solving (\ref{eq.1.3}) and thus imposing the following condition:
\begin{eqnarray}
\label{eq.3.2}
\frac{b}{a}=1+\frac{\Gamma_{1 1}^{1 0}}{\Gamma_{0 0}^{1 0}}>1
\end{eqnarray}
and the  reaction rates fulfill:
\begin{eqnarray}
\label{eq.3.3}
\Gamma_{1 1}^{1 0}=\Gamma_{1 1}^{0 1}>0;\; 2\Gamma_{0 0}^{1 0}= 2\Gamma_{0 0}^{0 1}\geq
\Gamma_{1 0}^{1 1}=\Gamma_{0 1}^{1 1}>0;\; \text{ and}\;
\Gamma_{1 0}^{0 0}=\Gamma_{0 1}^{0 0}=\frac{\Gamma_{1 1}^{1 0}(2\Gamma_{0 0}^{1 0}-\Gamma_{1 0}^{1 1})}{\Gamma_{0 0}^{1 0}}\geq 0
\end{eqnarray}
We will see that the case treated in this section (with the constraints (\ref{eq.3.3})) can be obtained from the reversible model of diffusion-coagulation with input of particles (RDCI model), solved in the next section, via a similarity transformation. In fact, in section VII, we investigate for a  local similarity transformation which would map the  general $S_{x,y}(t)$ function onto the empty-interval function (with $a=b$) and the RDCI model onto the present BCBD model. In the sequel we show that such a mapping exists and establish that the approaches followed in this section and in section VII are {\it equivalent}.

For the model (\ref{eq.3.1}) with the restriction (\ref{eq.3.3}) and from the definitions (\ref{eq.2.1}), we have $\alpha=\frac{2a\Gamma_{1 1}^{1 0}}{\Gamma_{0 0}^{1 0}}(\Gamma_{0 0}^{1 0}-\Gamma_{1 0}^{1 1})$, $\beta=-\frac{2}{a}(\Gamma_{0 0}^{1 0}-\Gamma_{1 0}^{1 1})$  and thus $\alpha\beta<0$,  $\mu_{1}=\mu_{2}=\mu=-i(sgn \alpha)\left|\frac{\alpha}{\beta}\right|^{1/2}$ and $q=i|\alpha\beta|^{1/2}$. We also have $\delta=2bA/a>0$ and $\gamma=2(B+C)-\delta$ (Note that because of (\ref{eq.3.3}), $0<|q|/\delta<1/2$).

 The subcase $\Gamma_{1 0}^{1 1}=\Gamma_{0 0}^{1 0}$ implies $\alpha=\beta=B=D=0$ and we recover (for $C\neq 0$) $\langle n_{x}(t)\rangle =\frac{a-S_{x,x+1}(t)}{b}=\frac{A}{C}+\left(\langle n_{x}(0)\rangle-\frac{A}{C}\right)e^{-2Ct}$.

 Hereafter we focus on the more general  situation where (\ref{eq.3.3}) are fulfilled with   $\Gamma_{1 0}^{1 1}\neq \Gamma_{0 0}^{1 0}$, and thus $\alpha\neq 0, \beta\neq 0$.

The stationary expression of the string-function for this model is given by the expression 
(\ref{eq.2.5}). With help of the formula (\ref{eq.2.15}) and the ratio (\ref{eq.3.2}), we obtain the following expression for the stationary density of particles in the system:
\begin{eqnarray}
\label{eq.3.4}
\langle n_{x}(\infty)\rangle=\frac{1}{b}\left(a-\mu\left[ \frac{J_{L+\gamma/\delta}(2q/\delta)Y_{1+\gamma/\delta}(2q/\delta)- Y_{L+\gamma/\delta}(2q/\delta)J_{1+\gamma/\delta}(2q/\delta) }
{J_{L+\gamma/\delta}(2q/\delta)Y_{\gamma/\delta}(2q/\delta)- Y_{L+\gamma/\delta}(2q/\delta)J_{\gamma/\delta}(2q/\delta)} \right]\right)
\end{eqnarray}

Similarly, with help of (\ref{eq.2.15}) and (\ref{eq.3.2}), we obtain the stationary expression of the instantaneous nearest-neighbour correlation functions:

\begin{eqnarray}
\label{eq.3.5}
\langle n_{x}n_{x+1}\rangle(\infty)=\frac{2ab\langle n_{x}(\infty)\rangle-a^{2}}{b^{2}}+(\frac{\mu}{b})^{2}\left[ \frac{J_{L+\gamma/\delta}(2q/\delta)Y_{2+\gamma/\delta}(2q/\delta)- Y_{L+\gamma/\delta}(2q/\delta)J_{2+\gamma/\delta}(2q/\delta) }
{J_{L+\gamma/\delta}(2q/\delta)Y_{\gamma/\delta}(2q/\delta)- Y_{L+\gamma/\delta}(2q/\delta)J_{\gamma/\delta}(2q/\delta)} \right]
\end{eqnarray}

From (\ref{eq.3.4}) and (\ref{eq.3.5}) one can check that the system under consideration is a 
{\it correlated system of interacting particles}. In fact, one can see that
$\langle n_{x}n_{x+1}\rangle (\infty)\neq\langle n_{x}(\infty)\rangle\langle n_{x+1}(\infty)\rangle=(\langle n_{x}(\infty)\rangle)^{2} $, despite the fact that both steady-states (\ref{eq.3.4}) and  (\ref{eq.3.5}) are {\it translationally-invariant}, the stationary distribution is {\it correlated} which is due to the interacting character
 ({\it hard-core}) of the particles. We should emphasize that the presence of  {\it correlations} in the stationary distribution is specific to
 the class of models considered here which  cannot be mapped onto {\it free-fermion} systems\footnote{ In fact, for the {\it free-fermion} systems such as the diffusion-limited with pair annihilation and creation model \cite{free-fermions} and the related diffusion-coagulation models \cite{Doering,benA,Hinrichsen}, it has been shown, for translationally-invariant systems, that $\langle n_{x}n_{x+r}\rangle(\infty)=(
\langle n_{x}(\infty)\rangle)^{2}\;,\forall r>0$} (see also the model of the next section).

To study the dynamical properties of the model, we need the {\it relaxation spectrum} of the string-function. As established in the previous section, the latter is obtained as the set of zeroes of the following Lommel function: $R_{L-1,1-E}(2i|q|/\delta)=0$, where $E\equiv \frac{q\lambda - \gamma}{\delta}$. Solving the associate eigenvalue-problem (\ref{eq.2.11}) (in this case, the matrix ${\cal M}$, see (\ref{eq.2.12}), is {\it not hermitian}).

For {\it small} systems the $(L-1)$  {\it distinct} eigenvalues $\{E_{i}\}$ of (\ref{eq.2.12}) can be computed analytically. For $L=6$, we have $\{E\}=\left\{3,3\pm\sqrt{\frac{5+4V^{2}\pm\sqrt{9+24V^{2}+4V^{4}}}{2}}\right\}$.
 For larger matrices we had to proceed numerically. Our analysis (based on the spectrum of large  matrices, with $L\leq 1000$), shows that the spectrum $\{E\}$ (and therefore $\{q\lambda\}$) is {\it always real} and symmetric  around $L/2$ which is an eigenvalue when $L$ is even. The other {\it eigenvalues} are not  generally {\it integers}, but for the {\it central} part of the spectrum (when eigenvalues which are close of  $L/2$), the eigenvalues approach integer values. This is not the case at the extremities of the spectrum. In particular, the smallest eigenvalue $E^{\ast}=min_{E}\{E\}$ is not an integer and depends on the size of the system:
$E^{\ast}=\epsilon_{L}>1$. However, for $L\gg 1$, $\epsilon_{L}\rightarrow \epsilon_{\infty}$, and  $E^{\ast}$ is a constant: $E^{\ast}=\epsilon_{\infty}>1$. For $L=6$, we have the exact result $\epsilon_{L=6}=3-\sqrt{\frac{5+4V^{2}+\sqrt{9+24V^{2}+4V^{4}}}{2}}$, with $1<\epsilon_{L=6}<3-\sqrt{2+\frac{1}{4}\sqrt{13}}$. This expression can be considered as an excellent approximation for systems of size  $L\gg 1$ and in particular for $\epsilon_{\infty}$ \footnote{As an illustration, for the case   $\Gamma_{0 0}^{1 0}=3/10$, $\Gamma_{1 0}^{1 1}=1/2$, $\Gamma_{1 1}^{1 0}=1$ and  $\Gamma_{1 0}^{0 0}=1/3$, with the expression above, we obtain (analytically) $\epsilon_{L=6}=1.0823337683$. For larger systems ($L= 10, 25, 40, 1000$), we obtain numerically (with an accuracy of $10^{-10}$): $\epsilon_{10}=\epsilon_{25}=\epsilon_{40}=\epsilon_{1000}=1.0823337697$.}.
 
 Therefore, the long-time dynamics (of large systems, with $L\gg 1$) is governed by the eigenvalue $E^{\ast}=\epsilon_{L}\simeq3-\sqrt{\frac{5+4V^{2}+\sqrt{9+24V^{2}+4V^{4}}}{2}} $ , i.e.
\begin{eqnarray}
\label{eq.3.6.0}
q\lambda^{\ast}=E^{\ast}\delta+\gamma =\epsilon_{L}\delta+\gamma
=2\left[\frac{\Gamma_{0 0}^{1 0}\Gamma_{1 0}^{1 1}+\Gamma_{1 1}^{1 0}(2\Gamma_{0 0}^{1 0}-
\Gamma_{1 0}^{1 1})}{\Gamma_{0 0}^{1 0}} + (\epsilon_{L}-1)(\Gamma_{0 0}^{1 0}+\Gamma_{1 1}^{1 0})\right]>2\Gamma_{1 0}^{1 1}\geq 0
\end{eqnarray}

The equation (\ref{eq.3.6.0}) provides the {\it inverse of the relaxation-time} of the system.

The dynamical approach towards the steady-state of the density is obtained from the dynamical expression (\ref{eq.2.14}) of the string-function, according to formula (\ref{eq.2.15}) 
and with the ratio (\ref{eq.3.2}), we obtain:
\begin{eqnarray}
\label{eq.3.7}
\langle n_{x}(t)\rangle-\langle n_{x}(\infty)\rangle=
\left(\frac{\mu}{b}\right)\sum_{E_{i}}{\cal A}_{E_{i}}e^{-(E_{i}\delta+\gamma)t}\left[J_{1-E_{i}}(2q/\delta)Y_{L-E_{i}}(2q/\delta) - Y_{1-E_{i}}(2q/\delta)J_{L-E_{i}}(2q/\delta)\right],
\end{eqnarray}
where the coefficients ${\cal A}_{E_{i}}$ have been computed for the translationally-invariant situation in (\ref{eq.2.15.0})-(\ref{eq.2.20.0}).

With help of  (\ref{eq.2.14}) and  (\ref{eq.2.15}) , we obtain the expression of the dynamical approach of the instantaneous nearest neighbour to its steady-state (\ref{eq.3.5}):
\begin{eqnarray}
\label{eq.3.8}
\langle n_{x} n_{x+1}\rangle(t)&-&\langle n_{x}n_{x+1}\rangle(\infty)=
\frac{a}{b}\left\{\langle n_{x}(t)\rangle+\langle n_{x+1}(t)\rangle -2\langle n_{x}(\infty)\rangle\right\}\nonumber\\&+&
\left(\frac{\mu}{b}\right)^{2}\sum_{E_{i}}{\cal A}_{E_{i}}e^{-(E_{i}\delta+\gamma)t}\left[J_{2-E_{i}}(2q/\delta)Y_{L-E_{i}}(2q/\delta) - Y_{2-E_{i}}(2q/\delta)J_{L-E_{i}}(2q/\delta)\right]
\end{eqnarray}
From (\ref{eq.2.14}), (\ref{eq.3.7}), (\ref{eq.3.8}) and with help of (\ref{eq.2.17.2})-(\ref{eq.2.17.5}), we can also obtain the {\it approximative} expression of the other two-point correlation functions.  

The result $(\ref{eq.3.7})$ can be extended to the case of the non-instantaneous two-point correlation functions $\langle n_{x}(t)n_{x_{0}}(0)\rangle$.
 It suffices to replace in (\ref{eq.3.7}) the coefficients ${\cal A}_{E_{i}}$ by those obtained from the initial condition $\left\langle \{\prod_{j=x}^{y-1}(a-bn_{j}(0))\}  n_{x_{0}}(0)\right\rangle$, instead of $S_{y-x}(0)$.

Let us now mention the long-time behaviour of quantities computed above. For the sake of simplicity we consider the translationally-invariant situation, in the regime where $E^{\ast}t\gg 1$ ($E^{\ast}$ is the  smallest  eigenvalue (\ref{eq.3.6.0})), we have:
\begin{eqnarray}
\label{eq.3.11}
\langle n_{x}(t)\rangle -\langle n_{x}(\infty)\rangle &\sim&
\left(\frac{\mu}{b}\right){\cal A}_{E^{\ast}}e^{-(E^{\ast}\delta+\gamma)t}\left[J_{1-E^{\ast}}(2q/\delta)Y_{L-E^{\ast}}(2q/\delta) - Y_{1-E^{\ast}}(2q/\delta)J_{L-E^{\ast}}(2q/\delta)\right]
\end{eqnarray}
\begin{eqnarray}
\label{eq.3.12}
\langle n_{x} n_{x+1}\rangle(t) -\langle n_{x} n_{x+1}\rangle (\infty)&\sim&
\frac{\mu^{2}{\cal A}_{E^{\ast}}}{b^{2}}e^{-(E^{\ast}\delta +\gamma)t}
\left[J_{2-E^{\ast}}\left(\frac{2q}{\delta}\right)Y_{L-E^{\ast}}\left(\frac{2q}{\delta}\right) -
Y_{2-E^{\ast}}\left(\frac{2q}{\delta}\right)J_{L-E^{\ast}}\left(\frac{2q}{\delta}\right)\right]
\nonumber\\&+& \frac{2a\mu{\cal A}_{E^{\ast}}}{b^{2}}e^{-(E^{\ast}\delta +\gamma)t}\left[
J_{1-E^{\ast}}\left(\frac{2q}{\delta}\right)Y_{L-E^{\ast}}\left(\frac{2q}{\delta}\right) -
Y_{1-E^{\ast}}\left(\frac{2q}{\delta}\right)J_{L-E^{\ast}}\left(\frac{2q}{\delta}\right)\right]
\end{eqnarray}

It follows from the exact results (\ref{eq.3.6.0})-(\ref{eq.3.12}), that the density $\langle n_{x}(t)\rangle$ and the two-point correlation functions $\langle n_{x}(t)n_{x_{0}}(0)\rangle$ (non-instantaneous) and $\langle n_{x}n_{x+1} \rangle (t)$ (instantaneous) approach the steady-state exponentially fast, as the string-function $S_{x,y}(t)$, with an inverse of relaxation-time given by (\ref{eq.3.6.0}).
In addition, in the sense of the approximative scheme (\ref{eq.2.17.2})-(\ref{eq.2.17.5}), these results are expected also to be valid for the more general correlation functions: $\langle n_{x} n_{x+r} \rangle(t) \;(r>1)$ and for arbitrary initial conditions: the initial state only  affects the multiplicative factors ${\cal A}_{E}$ of (\ref{eq.3.7}) and (\ref{eq.3.8}) (for translationally-invariant systems the ${\cal A}_{E}$ are given by (\ref{eq.2.15.0})). These statements  are supported by the study of the subcase $\Gamma_{1 0}^{1 1}=\Gamma_{0 0}^{1 0}$ (with $C\neq 0$) which is solvable by {\it conventional methods} and which dynamics is expected to be {\it qualitatively} the same as the more general case  considered here (especially for $|\Gamma_{1 0}^{1 1}-\Gamma_{0 0}^{1 0}|\ll 1$). For this  subcase the density has been computed previously (for arbitrary initial condition) and the two-point correlation functions read: $\langle n_{x}(t)n_{x_{0}}(0)\rangle=\frac{A}{C}+\left(\langle n_{x}n_{x_{0}}\rangle(0)-\frac{A}{C}\right)e^{-2Ct}$ and $\langle n_{x}n_{y}\rangle(t)=\left(\frac{A}{C}\right)^{2}(1-e^{-2Ct})+\langle n_{x}n_{y} \rangle(0) e^{-4Ct}+\frac{A}{C}\left(\langle n_{x}(0)\rangle + \langle n_{y}(0)\rangle -\frac{A}{C} \right)(e^{-2Ct}-e^{-4Ct})\;, (y\neq x)$.

\section{Solution of a reversible diffusion-coagulation model with input of particles}
In this section we study a  model of {\it reversible diffusion-coagulation with input of particles } which can be solved by the {\it conventional} IPDF method. Particles can jump (provided that the arrival site was previously empty) to the right and the left with rate $\Gamma_{1 0}^{0 1}=\Gamma_{0 1}^{1 0}>0$. We assume also that two adjacent particles can coagulate with the same rate $\Gamma_{1 1}^{1 0}=\Gamma_{1 1}^{0 1}$ and that when a particle is adjacent to a vacancy, a branching process can occur with rate  $\Gamma^{1 1}_{1 0}=\Gamma^{1 1}_{0 1}$. In addition, when two adjacent sites are empty, a particle can spontaneously appear on a site ({\it input}) with (a finite) rate $\Gamma_{0 0}^{1 0}=\Gamma_{0 0}^{0 1}>0$. The dynamics of this {\it reversible diffusion-coagulation with input of particles} (RDCI) model can be encoded by the following reactions:
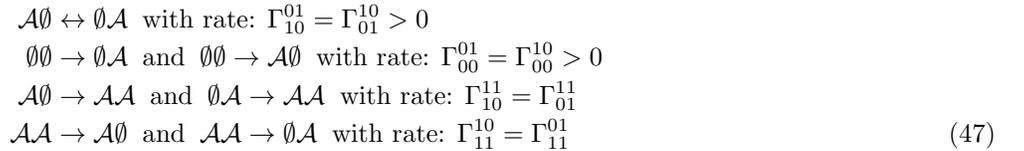
\begin{eqnarray}
\label{eq.4.1}
 {\cal A} \emptyset &\leftrightarrow& \emptyset {\cal A} \;\;\text{with rate: $\Gamma_{1 0}^{0 1}=\Gamma_{0 1}^{1 0} >0$} \nonumber\\
\emptyset \emptyset &\rightarrow& \emptyset {\cal A} \;\;\text{and} 
 \;\;\emptyset \emptyset \rightarrow  {\cal A} \emptyset \;\; \text{with rate: $\Gamma_{0 0}^{0 1}=\Gamma_{0 0}^{1 0}>0$} \nonumber\\
{\cal A} \emptyset &\rightarrow& {\cal A} {\cal A} \;\;\text{and} 
 \;\;\emptyset {\cal A} \rightarrow  {\cal A} {\cal A} \;\; \text{with rate: $\Gamma_{1 0}^{1 1}=\Gamma_{0 1}^{1 1}$} \nonumber\\
{\cal A} {\cal A}  &\rightarrow& {\cal A} \emptyset \;\;\text{and} 
 \;\; {\cal A}{\cal A}  \rightarrow  \emptyset {\cal A} \;\; \text{with rate:
 $\Gamma_{1 1}^{1 0}=\Gamma_{1 1}^{0 1}$}
\end{eqnarray}
This model can be solved by the {\it conventional} IPDF method: one can check that for the reactions (\ref{eq.4.1}) and (\ref{eq.4.2}) the constraints (\ref{eq.1.3}) are fulfilled with $a=b$.

In order to apply the ({\it conventional}) IPDF method, we require that the coagulation and diffusion rate are \cite{Doering,benA,Hinrichsen,Peschel}:
\begin{eqnarray}
\label{eq.4.2}
\Gamma_{1 0}^{0 1}=\Gamma_{1 1}^{0 1}\;\;\text{and} \;\;
\Gamma^{1 0}_{0 1}=\Gamma_{1 1}^{1 0}
\end{eqnarray}
Therefore with (\ref{eq.4.2}), we have $\Gamma_{1 0}^{0 1}=\Gamma_{1 1}^{0 1}=
\Gamma_{1 0}^{0 1}=\Gamma_{1 1}^{1 0}>0$ and are left with $4-1=3$ independent reaction-rates for the model (\ref{eq.4.1}).

Before proceeding with the solution of this model some comments are in order. To our knowledge this model has been studied in the continuum limit on an infinite chain (for the translationally invariant situation) by Doering and ben-Avraham \cite{Doering}. The latter obtained, in this limit, the  stationary concentration of particles, the stationary {\it interparticle function} and the relaxation spectrum as the zeroes of some Airy functions.
In their work, Doering  {\it et al.} considered an infinite chain with lattice spacing $\triangle x$ (here $\triangle x=1$ ) in the continuum limit ($\triangle x \rightarrow 0$) . On this infinite chain, the reactions occuring are the  symmetric diffusion and coagulation  with rates $\Gamma_{1 0}^{0 1}=\Gamma_{0 1}^{1 0}=
\Gamma_{1 1}^{0 1}=\Gamma_{1 1}^{1 0} =\frac{{\cal D}}{(\triangle x)^{2}}$, the symmetric branching processes with rate
$\Gamma_{1 0}^{1 1}=\Gamma_{0 1}^{1 1} =\frac{v}{\triangle x}$ and the {\it input} of particles with rate: $\Gamma_{0 0}^{1 0}=\Gamma_{0 0}^{0 1} =R\triangle x$. 

In this section we want to solve the model (\ref{eq.4.1}) with the restrictions (\ref{eq.4.2}) on a finite and periodic lattice of $L$ sites. According to the definitions (\ref{eq.2.1}),  for this section we have: $\alpha_{1}=\alpha_{2}\equiv\alpha\equiv 2(C_{1}-A_{1})=2\Gamma_{0 1}^{1 0}>0$, $\beta_{1}=\beta_{2}\equiv \beta \equiv-2D_{1}=2B_{1}=2(\Gamma_{1 0}^{0 1}+\Gamma_{1 0}^{1 1}-\Gamma_{0 0}^{1 0})$, 
 $\delta=A_{1}+A_{2}=2\Gamma_{0 0}^{1 0}>0$ and $\gamma+\delta=2(2\Gamma_{0 1}^{1 0}+\Gamma_{1 0}^{1 1})$.

Hereafter we thus solve the model (\ref{eq.4.1}) with the constraints  (\ref{eq.4.2}), which is a model described by $4-1=3$ independ paramaters (reaction-rates), for the case where the input of particles is non vanishing (i.e. $\delta>0$) and $D\neq 0$. The case where $\delta=0$ have been extensively studied in the continuum limit \cite{Doering,benA} as well as on discrete lattices \cite{Hinrichsen}: these models have been mapped onto {\it free-fermion} models \cite{Hinrichsen,Schutzrev,Henkel}.
In addition, when $D=0$, for the translationally-invariant situation and an initial density $\rho(0)$ of particles, we have (when $B\neq C$, see footnote  $(2)$)
\begin{eqnarray}
\label{eq.4.4.0}
\langle n_{x}(t)\rangle=\frac{A}{C-B}+\left(\rho(0)-\frac{A}{C-B}\right)e^{-2(C-B)t}.
\end{eqnarray}
For the model under consideration here, we have $\mu\equiv\sqrt{\frac{\alpha}{\beta}}$ and
 $q=\sqrt{\alpha \beta}$. As $\alpha>0$ and $\beta\geq 0$, with $\beta=0\Rightarrow D=B=0$, we focus on the case where $q\neq 0$. One should be cautious with the fact that $q$ and $\mu$ can be imaginary, in which case the sign ambiguity is fixed by requiring that $q\mu=\alpha$.

The equation of motion of the string-function associated to this model is thus of the form (\ref{eq.1.4}) which has been solved in section IV.

To compute the relevant physical quantities, we proceed as in the previous section. In fact the expressions of the density, correlation function and of the current of particles can be immediately obtained from  (\ref{eq.3.4})-(\ref{eq.3.11}) when $\frac{b}{a}=1$. Hereafter for the sake of completness and clarity we quote these expressions:

\begin{eqnarray}
\label{eq.4.4}
\langle n_{x}(\infty)\rangle=1-\mu\left[ \frac{J_{L+\gamma/\delta}(2q/\delta)Y_{1+\gamma/\delta}(2q/\delta)- Y_{L+\gamma/\delta}(2q/\delta)J_{1+\gamma/\delta}(2q/\delta) }
{J_{L+\gamma/\delta}(2q/\delta)Y_{\gamma/\delta}(2q/\delta)- Y_{L+\gamma/\delta}(2q/\delta)J_{\gamma/\delta}(2q/\delta)} \right]
\end{eqnarray}
\begin{eqnarray}
\label{eq.4.5}
\langle n_{x}n_{x+1}\rangle(\infty)=2\langle n_{x}(\infty)\rangle-1+\mu^{2}\left[ \frac{J_{L+\gamma/\delta}(2q/\delta)Y_{2+\gamma/\delta}(2q/\delta)- Y_{L+\gamma/\delta}(2q/\delta)J_{2+\gamma/\delta}(2q/\delta) }
{J_{L+\gamma/\delta}(2q/\delta)Y_{\gamma/\delta}(2q/\delta)- Y_{L+\gamma/\delta}(2q/\delta)J_{\gamma/\delta}(2q/\delta)} \right]
\end{eqnarray}
We again note that the expressions (\ref{eq.4.4}-\ref{eq.4.5}) are independent of the site label $x$ and therefore correspond to translationally-invariant stationary quantities.

As for the BCBD model, from (\ref{eq.4.4}) and (\ref{eq.4.5}) one can check that the model RDCI
is a {\it correlated system of interacting particles}, characterized by a translationally-invariant but {\it correlated} stationary distribution:
$\langle n_{x}n_{x+1}\rangle (\infty)\neq\langle n_{x}(\infty)\rangle\langle n_{x+1}(\infty)\rangle=(\langle n_{x}(\infty)\rangle)^{2} $.

To study the dynamical properties of the model, we need the {\it relaxation spectrum} $\{E_{i}\}, i=1,\dots,L$ of the {\it string function}. As established in the previous section, the latter is obtained as the set of zeroes of the following Lommel function: $R_{L-1,1-E}(2q/\delta)=0$, where $E\equiv\frac{q\lambda-\gamma}{\delta}$, which are computed solving the associated eigenvalue-problem (\ref{eq.2.11}). Notice that when $q$ is real the matrix ${\cal M}$ (\ref{eq.2.12}) is hermitian, otherwise ($q$ is imaginary) ${\cal M}$ is antihermitian.

Again the spectrum $\{E\}$ (and therefore $\{q\lambda\}$) turns out to be {\it real} (even when ${\cal M}$ is antihermitian) and symmetrically distributed  around $L/2$ which is an eigenvalue when $L$ is even.

When $\Gamma_{0 0}^{1 0}>\Gamma_{0 1}^{1 0}+\Gamma_{0 1}^{1 1}$, then $q$ is imaginary, with $0<\frac{|q|}{\delta}\leq \frac{1}{2}$.  When $q$ is imaginary, the  {\it eigenvalues} are not  generally {\it integers}, but for the {\it central} part of the spectrum (when eigenvalues which are close of  $L/2$), the eigenvalues approach integer values. This is not the case for  the smallest eigenvalue $E^{\ast}=min_{E}\{E\}$ which is not an integer and depends on the size of the system:
$E^{\ast}=\epsilon_{L}>1$. However, for $L\gg 1$, 
$\epsilon_{L}\rightarrow \epsilon_{\infty}$, and  $E^{\ast}$ is a constant: $E^{\ast}=\epsilon_{\infty}>1$. The expression (\ref{eq.2.13.0}) can again be considered as an excellent approximation for systems of size  $L\gg 1$ and in particular for $\epsilon_{\infty}$. We can then show that for $q$ imaginary, the quantity $\epsilon_{L}>1$. In fact one can compute  (approximative) bounds for $\epsilon_{L}$, with $L \gg 1$.  This is achieved with help of (\ref{eq.2.13.0}), namely:
 $1<\epsilon_{L=6}\leq 3-\sqrt{2+\frac{1}{4}\sqrt{13}}$ . 

When $\Gamma_{0 0}^{1 0}<\Gamma_{0 1}^{1 0}+\Gamma_{0 1}^{1 1}$, $q$ is real. The eigenvalues are still symetrically distributed around $L/2$ (which is still an eigenvalue when $L$ is even). In this case, the smallest eigenvalue $E^{\ast}\equiv\epsilon_{L}$ can be negative (e.g., $\epsilon_{L=6}<0$ if $\frac{q}{\delta}>1.29907...$). However, $ -\frac{2\Gamma_{0 1}^{1 0}}{\Gamma_{0 0}^{1 0}} \lesssim \epsilon_{L}\lesssim 1$ (in fact, $-\frac{2\Gamma_{0 1}^{1 0}}{\Gamma_{0 0}^{1 0}}< \epsilon_{L=6}\leq 1$ ), and thus, when $q$ is real, we have:
$q\lambda^{\ast}=\epsilon_{L}\delta+\gamma=
2\left[2\Gamma^{1 0}_{0 1}+\Gamma_{1 0}^{1 1}+(\epsilon_{L}-1)\Gamma_{0 0}^{1 0}\right]
\geq 2(\Gamma_{0 1}^{1 0}+\epsilon_{L}\Gamma_{0 0}^{1 0})>0 $.
 Again, for large systems ($L\gg 1$) the expression of  $\epsilon_{L}$ is well approximated by the exact expression (\ref{eq.2.13.0}) of $\epsilon_{L=6}$.

When  $\Gamma_{0 0}^{1 0}=\Gamma_{0 1}^{1 0}+\Gamma_{0 1}^{1 1}$, then (with (\ref{eq.4.2})) $D=B=q=0$ and we recover (\ref{eq.4.4.0}), with $B=0$). 

In definitive, the long-time dynamics of large systems is governed by the eigenvalue $E^{\ast}>0 $ according to (\ref{eq.2.13.0}) for $L\gg 1$. This means that the {\it inverse of the relaxation-time} of the system reads:
\begin{eqnarray}
\label{eq.4.6.0}
q\lambda^{\ast}=E^{\ast}\delta+\gamma =\epsilon_{L}\delta+\gamma
=2\left[2\Gamma^{1 0}_{0 1}+\Gamma_{1 0}^{1 1}+(\epsilon_{L}-1)\Gamma_{0 0}^{1 0}\right]>0
\end{eqnarray}

The dynamical part of the above quantities are obtained similarly from (\ref{eq.3.7})-(\ref{eq.3.8}), setting $\frac{b}{a}=1$ in these expressions:
\begin{eqnarray}
\label{eq.4.7}
\langle n_{x}(t)\rangle-\langle n_{x}(\infty)\rangle=
\mu\sum_{E_{i}}{\cal A}_{E_{i}}e^{-(E_{i}\delta+\gamma)t}\left[J_{1-E_{i}}(2q/\delta)Y_{L-E_{i}}(2q/\delta) - Y_{1-E_{i}}(2q/\delta)J_{L-E_{i}}(2q/\delta)\right]
\end{eqnarray}
\begin{eqnarray}
\label{eq.4.8}
\langle n_{x} n_{x+1}\rangle(t)&-&\langle n_{x}n_{x+1}\rangle(\infty)=
\left\{\langle n_{x}(t)\rangle+\langle n_{x+1}(t)\rangle -2\langle n_{x}(\infty)\rangle\right\}\nonumber\\&+&
\mu^{2}\sum_{E_{i}}{\cal A}_{E_{i}}e^{-(E_{i}\delta+\gamma)t}\left[J_{2-E_{i}}(2q/\delta)Y_{L-E_{i}}(2q/\delta) - Y_{2-E_{i}}(2q/\delta)J_{L-E_{i}}(2q/\delta)\right],
\end{eqnarray}
where the coefficients ${\cal A}_{E_{i}}$ have been computed in the translationally-invariant situation in (\ref{eq.2.15.0})-(\ref{eq.2.20.0}).

From equations (\ref{eq.4.7}),(\ref{eq.4.8}) and (\ref{eq.2.14}), we can compute for this RDCI model the approximative expression all two-point correlation functions according to the scheme (\ref{eq.2.17.2})-(\ref{eq.2.17.5}).  

The long-time behaviour of the quantities (\ref{eq.4.7})-(\ref{eq.4.8}) are also obtained as explained in the previous section and in particular the long-time behaviour of the density is given by (\ref{eq.3.11}), where the smallest eigenvalue $E^{\ast}$ is the quantity obtained in (\ref{eq.4.6.0}).

As in the case of the model BCBD, and for the same reasons, the density and the two-point correlation functions (instantaneous and non-instantaneous) relaxe exponentially fast to the steady-state with an inverse of relaxation-time given by (\ref{eq.4.6.0}).

Another relevant quantity which one can compute for this model is the 
so-called {\it interparticle function} $p_{x,y}(t)$ \cite{Doering,benA}. The latter give the probability, a time $t$, for a particle at site $x-1$ to have as a {\it next neighbour}  particle at a distant site $y-x>0$.
To obtain $p_{x,y}(t)$, we set $a=b=1$ and the string-function $S_{x,y}(t)$ reduces to the {\it empty-interval function} \cite{Doering,benA}: $ S_{x,y}(t)=\langle\prod_{j=x}^{y-1}(1-n_{j})\rangle (t)$ which is associated to the probability of having a {\it sequence of holes} starting from the site $x$ and of length $y-x$. As shown by Doering et {\it al.} \cite{Doering,benA}, it is possible to relate the density of particles, the {\it empty-interval function} $S_{x,y}$ and the {\it interparticle function} $p_{x,y}(t): S_{x,y+1}(t)-2S_{x,y}(t)+S_{x,y-1}(t)=\left\langle (1-n_{x})\dots (1-n_{y-2}) \left(n_{y-1}-(1-n_{y-1})n_{y}\right)\right\rangle(t)$.
Therefore,  with $\rho(t)=\frac{1}{L}\sum_{j}\langle n_{j}(t)\rangle$,  for the translationally-invariant situation (the inverse of the $\rho(t)$ measures the average distance between adjacent particles), we have:
\begin{eqnarray}
\label{eq.4.11}
p_{x,y}(t)=p_{y-x}(t)= \frac{\left[S_{x,y+1}(t)-2S_{x,y}(t)+S_{x,y-1}(t)\right]}{\rho(t)}
\end{eqnarray}

 In particular in the stationary case, we have:
\begin{eqnarray}
\label{eq.4.12}
&&p_{x,y}(\infty)=\frac{S_{x,y+1}(\infty)-2S_{x,y}(\infty)+S_{x,y-1}(\infty)}{\rho(\infty)}
\nonumber\\
&=&\mu^{y-x}\left(\frac{
J_{L+\frac{\gamma}{\delta}}(\frac{2q}{\delta})
\left[ \mu Y_{y-x+1+\frac{\gamma}{\delta}}(\frac{2q}{\delta})
+\mu^{-1} Y_{y-x-1+\frac{\gamma}{\delta}}(\frac{2q}{\delta})
-2 Y_{y-x+\frac{\gamma}{\delta}}(\frac{2q}{\delta})
\right]
}
{J_{L+\frac{\gamma}{\delta}}(\frac{2q}{\delta})
[Y_{\frac{\gamma}{\delta}}(\frac{2q}{\delta})- \mu Y_{1+\frac{\gamma}{\delta}}(\frac{2q}{\delta}) ]
-
Y_{L+\frac{\gamma}{\delta}}(\frac{2q}{\delta})
[J_{\frac{\gamma}{\delta}}(\frac{2q}{\delta})- \mu J_{1+\frac{\gamma}{\delta}}(\frac{2q}{\delta}) ]} \right)
\nonumber\\
 &-&
\mu^{y-x}\left(\frac{
Y_{L+\frac{\gamma}{\delta}}(\frac{2q}{\delta})
\left[ \mu J_{y-x+1+\frac{\gamma}{\delta}}(\frac{2q}{\delta})
+\mu^{-1} J_{y-x-1+\frac{\gamma}{\delta}}(\frac{2q}{\delta})
-2 J_{y-x+\frac{\gamma}{\delta}}(\frac{2q}{\delta})
\right]
}
{J_{L+\frac{\gamma}{\delta}}(\frac{2q}{\delta})
[Y_{\frac{\gamma}{\delta}}(\frac{2q}{\delta})- \mu Y_{1+\frac{\gamma}{\delta}}(\frac{2q}{\delta}) ]
-
Y_{L+\frac{\gamma}{\delta}}(\frac{2q}{\delta})
[J_{\frac{\gamma}{\delta}}(\frac{2q}{\delta})- \mu J_{1+\frac{\gamma}{\delta}}(\frac{2q}{\delta}) ]} \right)
\end{eqnarray}

To conclude this section, let us point out the fact that the results (\ref{eq.4.4})-(\ref{eq.4.12}) can be immediately generalized to systems which, in addition to the processes (\ref{eq.4.1}), also include the (adjacent) pair-creation reaction: $\emptyset\emptyset \longrightarrow {\cal A}{\cal A}$, with rate $\Gamma_{0 0}^{1 1}$. In fact, it suffices to replace $\Gamma_{0 0}^{1 0}$ with
  $\Gamma_{0 0}^{1 0}+\Gamma_{0 0}^{1 1} $ in  (\ref{eq.4.4})-(\ref{eq.4.12}).

We also would like to emphasize the fact that the expressions  (\ref{eq.4.4})-(\ref{eq.4.12}) are different from those obtained by Doering and ben-Avraham \cite{Doering} who considered the continuum limit of this model on an infinite chain.

\section{Solution, via similarity transformations, of models which cannot be mapped onto free-fermion systems}

In sections V and VI we have solved two different reaction-diffusion, (the BCBD and the RDCI), models which cannot be mapped onto free-fermion systems.  It is therefore natural to wonder whether or not there exists a  mapping between these models, transforming the {\it empty-interval function} $\langle\prod_{j=x}^{y-1}(1-n_{j})\rangle (t)$ into another {\it string-function} of the form $\langle\prod_{j=x}^{y-1}(1-\frac{b}{a}n_{j})\rangle (t)$.
 In this section we  study the class of  models which can be obtained from the model of reversible diffusion-coagulation with input of particles (RDCI), analyzed in section VI through a class of local {\it similarity transformations}. In so doing we will answer the following questions:

(i) Does  a {\it similarity transformation} exist which maps the RDCI onto the BCBD model and the {\it empty-interval function} (with $a=b$) onto a {\it string-function} (with $\frac{b}{a}\neq 1$) ?

(ii)  
If so, does the mapping provide the solution of the model BCBD for the same constraints (\ref{eq.3.3}) considered in section V ?

Let us consider the {\it original} stochastic Hamiltonian $H$, through the (local) similarity 
transformation ${\cal B}$, define  $\widetilde{H}$  as \cite{Hinrichsen,Schutzrev,Simon}
\begin{eqnarray}
\label{eq.5.1}
\widetilde{H}\equiv {\cal B}^{-1}H{\cal B}
\end{eqnarray}
Because of the requirement that $\langle \widetilde{O}(t)\rangle(\widetilde{H},|\widetilde{P}(0)\rangle)=\langle O(t)\rangle(H,|P(0)\rangle) $, which implies that  $\langle\widetilde{\chi}|Oe^{-Ht}|P(0)\rangle
=\langle\widetilde{\chi}|\widetilde{O}e^{-\widetilde{H}t}|\widetilde{P}(0)\rangle=\langle\widetilde{\chi}|\widetilde{O}{\cal B}^{-1}e^{-Ht}{\cal B}|\widetilde{P}(0)\rangle=\langle\widetilde{\chi}|O {\cal B} {\cal B}^{-1}e^{-Ht} {\cal B}{\cal B}^{-1}|P(0)\rangle$, it is clear that under this transformation, the observable $O$, and the initial state $|P(0)\rangle$ transform according to:
$
\widetilde{O}\equiv O{\cal B}\;\; \text{and} \;\;|\widetilde{P}(0)\rangle\equiv{\cal B}^{-1} |P(0)\rangle ,
$
where we assume homogeneous (uncorrelated, yet random) initial states with density $0\leq \rho(0)\leq 1$ of particles: $|P(0)\rangle =
\left(
 \begin{array}{c}
 1-\rho(0)\\
\rho(0)
 \end{array}\right)$.
In this section we focus on local transformations of the form \cite{Schutzrev,Hinrichsen,Simon}
\begin{eqnarray}
\label{eq.5.3}
{\cal B}=\bigotimes_{j=1}^{L}{\bf b}_{j},
\end{eqnarray}
where ${\bf b}_{j}$ denotes a $2\times 2$ matrix acting at site $j$ such that the stochasticity condition $\langle \widetilde{\chi}|{\cal B}\widetilde{H}=0$ is fulfilled.
In addition, in order to transform the {\it empty-interval function} into a more general {\it string-function} we want to consider transformations which map the operator $1-n_{j}$
onto $\widetilde{1-n_{j}}\equiv 1-rn_{j}$, where  
$r\equiv\frac{b}{a}$. Peschel et {\it al.} \cite{Peschel} have shown that the transformation  (\ref{eq.5.3}), with 
\begin{eqnarray}
\label{eq.5.4}
{\bf b}_{j}\equiv
\left(
 \begin{array}{c c }
1 & 1-r  \\
 0  & r 
 \end{array}\right)_{j}\,
\end{eqnarray}
satisfy this property.
Through this transformation, according to (\ref{eq.5.1}), the stochastic Hamiltonian related to the RDCI model transforms into the following stochastic Hamiltonian
$
-\widetilde{H}_{j, j+1} =
\left(
 \begin{array}{c c c c}
 \widetilde{\Gamma}_{0 0}^{0 0} & \widetilde{\Gamma}_{0 1}^{0 0} &\widetilde{\Gamma}_{1 0}^{0 0} &\widetilde{\Gamma}_{1 1}^{0 0}  \\
 \widetilde{\Gamma}_{0 0}^{0 1} & \widetilde{\Gamma}_{0 1}^{0 1} &\widetilde{\Gamma}_{1 0}^{0 1} &\widetilde{\Gamma}_{1 1}^{0 1} \\
 \widetilde{\Gamma}_{0 0}^{1 0} & \widetilde{\Gamma}_{0 1}^{1 0} &\widetilde{\Gamma}_{1 0}^{1 0} &\widetilde{\Gamma}_{1 1}^{1 0} \\
 \widetilde{\Gamma}_{0 0}^{1 1} & \widetilde{\Gamma}_{0 1}^{1 1}&\widetilde{\Gamma}_{1 0}^{1 1} &\widetilde{\Gamma}_{1 1}^{1 1} \\
 \end{array}\right)\,
$
where the non-diagonal entries read:
\begin{eqnarray}
\label{eq.5.6}
\widetilde{\Gamma}_{0 0}^{0 1}&=& \widetilde{\Gamma}_{0 0}^{1 0}/r=\Gamma_{0 0}^{1 0}/r \;\;;\;\;
\widetilde{\Gamma}_{0 0}^{1 1}= 0\;\;;\;\;
\widetilde{\Gamma}_{0 1}^{0 0}=\widetilde{\Gamma}_{1 0}^{0 0}= 
 (2\Gamma_{0 0}^{1 0}-\Gamma_{1 0}^{1 1})(r-1)/r
\nonumber\\
\widetilde{\Gamma}_{0 1}^{1 0}&=& \widetilde{\Gamma}^{0 1}_{1 0}=
[(\Gamma_{1 0}^{1 1}-\Gamma_{0 0}^{1 0})(r-1)+r\Gamma_{0 1}^{1 0}]/r \;\;;\;\;
\widetilde{\Gamma}_{0 1}^{1 1}=\widetilde{\Gamma}_{1 0}^{1 1}=\Gamma_{1 0}^{1 1}/r\;\;;\;\;
\widetilde{\Gamma}_{1 1}^{0 0}=2(r-1)[(r-1)(\Gamma_{1 0}^{1 1}-\Gamma_{0 0}^{1 0})+r\Gamma_{0 1}^{1 0}]/r
 \nonumber\\
\widetilde{\Gamma}_{1 1}^{0 1} &=&\widetilde{\Gamma}_{1 1}^{1 0} =
[\Gamma_{0 0}^{1 0}(r-1)^{2}+(2-r)((r-1)\Gamma_{1 0}^{1 1}+r\Gamma_{0 1}^{1 0})]/r
\end{eqnarray}
The uncorrelated and homogeneous, but random, initial state becomes:
\begin{eqnarray}
\label{eq.5.7}
|\widetilde{P}(0)\rangle =
\left(
 \begin{array}{c}
 1-\frac{\rho(0)}{r}\\
\frac{\rho(0)}{r}
 \end{array}\right)\,
\end{eqnarray}
One has to ensure that all the reaction rates appearing in (\ref{eq.5.6}) are $\geq 0$, which require that $r\geq 0$ and that $0\leq\frac{\rho(0)}{r}\leq 1$. Therefore we have the necessary condition: $r\geq \rho(0)\geq 0$.

According to the transformation (\ref{eq.5.3}), (\ref{eq.5.4}) the {\it empty-interval function} $\langle \prod_{j=x}^{y-1}(1-n_{j})\rangle(t)$ is mapped onto the {\it string-function}:
 $\langle \prod_{j=x}^{y-1}\widetilde{(1-n_{j})}\rangle(t)=
\langle \prod_{j=x}^{y-1}(1-rn_{j})\rangle(t)$ and therefore, the statistical quantities for the model described by the stochastic Hamiltonian $\widetilde{H}$ are obtained from the related quantities computed in section VI for the model RDCI as
\begin{eqnarray}
\label{eq.5.8}
\langle n_{x}(t)\rangle_{\widetilde{H},|\widetilde{P}(0)\rangle} =\frac{1}{r}\langle n_{x}(t)\rangle_{H,|P(0)\rangle } \;\;;\;\;
\langle n_{x} n_{x+y} \rangle_{\widetilde{H},|\widetilde{P}(0)\rangle} (t)=\frac{1}{r^{2}}\langle n_{x}n_{x+y}\rangle_{H,|P(0)\rangle }(t)\;\;;(y>0)
\end{eqnarray}
where, $\langle n_{x}(t)\rangle_{H,|P(0)\rangle }$ and $\langle n_{x}n_{x+1}\rangle_{H,|P(0)\rangle }(t)$ have been computed in (\ref{eq.4.7}) and (\ref{eq.4.8}).

We will now consider two specific models described by  (\ref{eq.5.6}).

(a) To answer the questions (i) and (ii), we seek, through the mapping (\ref{eq.5.3}),(\ref{eq.5.4}), a model of the BCBD-type and thus require,  $\widetilde{\Gamma}_{1 0}^{0 1}=
\widetilde{\Gamma}_{0 1}^{1 0}=\widetilde{\Gamma}_{1 1}^{0 0}=0$, as for the
BCBD model considered in section V and infer from (\ref{eq.5.6}) : $(\Gamma_{1 0}^{1 1}-\Gamma_{0 0}^{1 0})(r-1)+r\Gamma_{0 1}^{1 0}=0$ ($r\neq 1$), which implies: 
\begin{eqnarray}
\label{eq.5.9}
r=\frac{\Gamma_{1 0}^{1 1}-\Gamma_{0 0}^{1 0}}{\Gamma_{1 0}^{1 1}+\Gamma_{0 1}^{1 0}-\Gamma_{0 0}^{1 0} }\geq 0
\end{eqnarray}

Replacing the expression (\ref{eq.5.9}) in (\ref{eq.5.6}), we obtain the reaction-rates of the new (BCBD) model in terms of the rates of the original (RDCI) model. In order to have a physical BCBD model, we have to require the reaction rates $\widetilde{\Gamma}$'s  to be positive. We now take advantage from the fact that a version of the BCBD model has been solved in section V, where ${\widetilde{\Gamma}}_{1 0}^{0 0}=\frac{\widetilde{\Gamma}_{1 1}^{1 0}}{\widetilde{\Gamma}_{0 0}^{1 0}}(2\widetilde{\Gamma}_{0 0}^{1 0}-\widetilde{\Gamma}_{1 0}^{1 1})$. It is therefore possible to {\it check}, from  (\ref{eq.5.6}) and (\ref{eq.5.9}), that this relation still holds in this case and parametrize the reaction rates  (\ref{eq.5.6}) as follows:
\begin{eqnarray}
\label{eq.5.10}
\widetilde{\Gamma}_{0 0}^{1 0}&=&\Gamma_{0 0}^{1 0}/r,\;\;
\widetilde{\Gamma}_{1 0}^{1 1}=\Gamma_{1 0}^{1 1}/r,\;\;
\widetilde{\Gamma}_{1 1}^{1 0}=(r-1)\widetilde{\Gamma}_{0 0}^{1 0},\;\; 
\widetilde{\Gamma}_{1 0}^{0 0}=\frac{\widetilde{\Gamma}_{1 1}^{1 0}}{\widetilde{\Gamma}_{0 0}^{1 0}}(2\widetilde{\Gamma}_{0 0}^{1 0}-\widetilde{\Gamma}_{1 0}^{1 1}),\;\; r=1+\frac{\widetilde{\Gamma}_{1 1}^{1 0}}{\widetilde{\Gamma}_{0 0}^{1 0}}>1
\end{eqnarray}
The requirement of positivity of these rates (\ref{eq.5.10}) leads to  $2\widetilde{\Gamma}_{0 0}^{1 0}>\widetilde{\Gamma}_{1 1}^{1 0}$. Thus, the reaction-rates (\ref{eq.5.10}) describing the BCBD model from the RDCI model, are identical to those considered in section V for solving the BCBD {\it directly} from the generalized string function. It is therefore easy to obtain the density and the correlation functions of the BCBD model from  the RDCI model inverting the relation (\ref{eq.5.10}) and using  (\ref{eq.5.7}), (\ref{eq.5.8}). As an illustration, we consider the BCBD model with rates $\widetilde{\Gamma}_{0 0}^{1 0}=1\;,\widetilde{\Gamma}_{1 0}^{1 1}=1/2\;,\widetilde{\Gamma}_{1 1}^{1 0}=2$ and $\widetilde{\Gamma}_{1 0}^{0 0}=3$, which imply that $r=3$. Such a model is thus mapped onto the RDCI model with the rates $\Gamma_{0 0}^{1 0}=3\;,\Gamma_{1 0}^{1 1}=3/2\;,\Gamma_{0 1}^{1 0}=\Gamma_{1 1}^{1 0}=1$. In this case the density of the BCBD model is related to the density of the RDCI model according to (\ref{eq.5.8}):
$\langle n_{x}(t)\rangle^{BCBD}_{(\widetilde{\Gamma}_{0 0}^{1 0}=1\;,\widetilde{\Gamma}_{1 0}^{1 1}=1/2\;,\widetilde{\Gamma}_{1 1}^{1 0}=2\;,\widetilde{\Gamma}_{1 0}^{0 0}=3;|\widetilde{P}(0)\rangle)}=
\frac{1}{3}\langle n_{x}(t)\rangle^{RDCI}_{(\Gamma_{0 0}^{1 0}=3\;,\Gamma_{1 0}^{1 1}=3/2\;,\Gamma_{0 1}^{1 0}=\Gamma_{1 1}^{1 0}=1;|P(0)\rangle)}$.
In particular we have seen that the stationary density is independent of the initial state and thus, in this case, we have (see (\ref{eq.3.4}) and (\ref{eq.4.4})):
$\langle n_{x}(\infty)\rangle^{BCBD}_{(\widetilde{\Gamma}_{0 0}^{1 0}=1\;,\widetilde{\Gamma}_{1 0}^{1 1}=1/2\;,\widetilde{\Gamma}_{1 1}^{1 0}=2\;,\widetilde{\Gamma}_{1 0}^{0 0}=3)}=
\frac{1}{3}\langle n_{x}(\infty)\rangle^{RDCI}_{(\Gamma_{0 0}^{1 0}=3\;,\Gamma_{1 0}^{1 1}=3/2\;,\Gamma_{0 1}^{1 0}=\Gamma_{1 1}^{1 0}=1)}=\frac{1}{3}+\frac{i\sqrt{2}}{3}\left[\frac{J_{L+\frac{1}{6}}(\sqrt{2}i/3)Y_{\frac{7}{6}}(\sqrt{2}i/3)-
Y_{L+\frac{1}{6}}(\sqrt{2}i/3)J_{\frac{7}{6}}(\sqrt{2}i/3)
 }{
J_{L+\frac{1}{6}}(\sqrt{2}i/3)Y_{\frac{1}{6}}(\sqrt{2}i/3)-
Y_{L+\frac{1}{6}}(\sqrt{2}i/3)J_{\frac{1}{6}}(\sqrt{2}i/3)
}\right]\simeq 0.2401 \;\;(L\gg 1).$

We are now in a position to answer the questions (i) and (ii):

{\it We have shown that there exists a similarity transformation (\ref{eq.5.3}), (\ref{eq.5.4}) that transforms the empty-interval function onto the generalized string-function and that maps the RDCI onto the BCBD model, with the same constraints of solvability that the constraints (\ref{eq.3.3}) imposed in section V. We conclude that the present approach and the method devised in section V are equivalent.}

 One additional comment on this equivalence is however useful at this point. Although both mentioned methods  are equivalent, the method devised in section V is in a sense {\it more convenient} because it is {\it direct}: solving the equation of motion of the adequate (generalized) string-function, solves directly the BCBD model. Conversely, via the similarity transformation, we first  solve the RDCI model, which is a task of the same difficulty as that  of solving the BCBD model, and then  find an adequate and non-trivial similarity transformation (where the new reaction-rates should be interpreted correctly in term of the original ones).

\vspace{0.5cm}
(b) Let us now consider a model which can be solved from the solution of the RDCI model via the similarity transformation  (\ref{eq.5.3}),(\ref{eq.5.4}).
The model under consideration is a reversible diffusion-coagulation with particles input and pair annihilation (RDCIPA), which dynamics can be symbolized by the reactions:
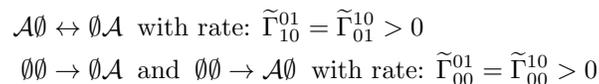
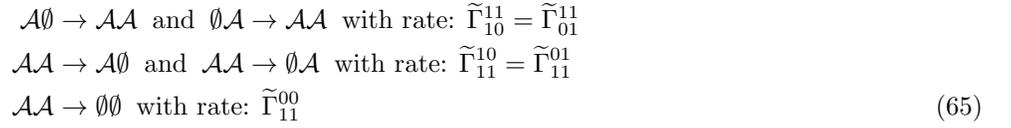
\begin{eqnarray}
\label{eq.5.11}
 {\cal A} \emptyset &\leftrightarrow& \emptyset {\cal A} \;\;\text{with rate: $\widetilde{\Gamma}_{1 0}^{0 1}=\widetilde{\Gamma}_{0 1}^{1 0} >0$} \nonumber\\
\emptyset \emptyset &\rightarrow& \emptyset {\cal A} \;\;\text{and} 
 \;\;\emptyset \emptyset \rightarrow  {\cal A} \emptyset \;\; \text{with rate: $\widetilde{\Gamma}_{0 0}^{0 1}=\widetilde{\Gamma}_{0 0}^{1 0}>0$} \nonumber\\
{\cal A} \emptyset &\rightarrow& {\cal A} {\cal A} \;\;\text{and} 
 \;\;\emptyset {\cal A} \rightarrow  {\cal A} {\cal A} \;\; \text{with rate: $\widetilde{\Gamma}_{1 0}^{1 1}=\widetilde{\Gamma}_{0 1}^{1 1}$} \nonumber\\
{\cal A} {\cal A}  &\rightarrow& {\cal A} \emptyset \;\;\text{and} 
 \;\; {\cal A}{\cal A}  \rightarrow  \emptyset {\cal A} \;\; \text{with rate:
 $\widetilde{\Gamma}_{1 1}^{1 0}=\widetilde{\Gamma}_{1 1}^{0 1}$}\nonumber\\
{\cal A} {\cal A}  &\rightarrow& \emptyset \emptyset  \;\; \text{with rate:
 $\widetilde{\Gamma}_{1 1}^{0 0}$}
\end{eqnarray}
This model(RDCIPA) can be obtained from the RDCI model via the similarity transformation (\ref{eq.5.3}), (\ref{eq.5.4}). Imposing $\Gamma_{1 0}^{1 1}=2\Gamma_{0 0}^{1 0}$ in (\ref{eq.5.6})  we get the following reaction-rates:
\begin{eqnarray}
\label{eq.5.12}
\widetilde{\Gamma}_{0 0}^{1 0}&=&\widetilde{\Gamma}_{0 0}^{0 1}= \Gamma_{0 0}^{1 0}/r,\;\;\widetilde{\Gamma}_{1 0}^{1 1}=\widetilde{\Gamma}_{0 1}^{1 1}=
2\Gamma_{0 0}^{1 0},\;\;\widetilde{\Gamma}_{1 1}^{0 0}=2(r-1)\;\widetilde{\Gamma}_{0 1}^{1 0}\nonumber\\
\widetilde{\Gamma}_{1 1}^{1 0}&=&\widetilde{\Gamma}_{1 1}^{0 1}=(2-r)\widetilde{\Gamma}_{0 1}^{1 0}+(r-1)\widetilde{\Gamma}_{0 0}^{1 0},\;\;
\widetilde{\Gamma}_{0 1}^{1 0}=\widetilde{\Gamma}_{0 1}^{1 0}=\frac{r-1}{2}\widetilde{\Gamma}_{1 0}^{1 1}+\Gamma_{1 0}^{0 1}
\end{eqnarray}
For this (RDCIPA) model, we have three (positive) independent parameters: $r\geq 1$, $\Gamma_{0 0}^{1 0}\geq 0$ and $\Gamma_{0 1}^{1 0}\geq 0$.
 The positivity of the reaction-rates (\ref{eq.5.11}) and the physical meaning of the initial state requires the following constraints:
\begin{eqnarray}
\label{eq.5.14}
\widetilde{\Gamma}_{1 1}^{0 1}=(2-r)\widetilde{\Gamma}_{0 1}^{1 0}+(r-1)\widetilde{\Gamma}_{0 0}^{1 0}\geq 0,\;\; r\geq 1,\;\;0\leq \rho(0)\leq r
\end{eqnarray}
Thus, for the model RDCIPA (\ref{eq.5.11}) described by the reaction-rates (\ref{eq.5.12}) with the restrictions  (\ref{eq.5.14}), the density, the correlation functions can be computed from the results of the model RDCI according to (\ref{eq.5.8}), for homogeneous (but random) initial states described by (\ref{eq.5.7}). 
\section{Propagation of a wave-front  and the Fisher waves}
At the end of section II, we have stated that some reaction-diffusion models, are described at the mean-field level and in the continuum limit  by non-linear partial differential equation of Fisher type (\ref{eq.0.11}) \cite{Fisher}. In this section we show that for some choices of the parameters (reaction-rates) the mean-field formulation of the models BCBD and RDCI gives rise to {\it Fisher-type} equations. Then, from the results obtained in section V and VI, we study the propagation of the wave-front from a microscopic point of view (in so doing, the correlation between particles are  taken into account {\it exactly}). We show that the scenario predicted by Fisher's theory fails in one spatial dimension for the models under consideration.
In all this section, we adopt the same notation as that introduced at the end of the section II.

(i) For the model BCBD, setting
\begin{eqnarray}
\label{eq.5.15}
\phi_{BCBD}\equiv \frac{(2\Gamma_{0 0}^{1 0}+\Gamma_{1 0}^{0 0}- \Gamma_{1 0}^{1 1})+\sqrt{
(\Gamma_{1 0}^{1 1})^{2}+(\Gamma_{1 0}^{0 0})^{2}+4\Gamma_{0 0}^{1 0}\Gamma_{1 1}^{1 0}-2\Gamma_{1 0}^{1 1}\Gamma_{1 0}^{0 0} }}{2\left[(\Gamma_{1 0}^{0 0}+\Gamma_{0 0}^{1 0})-(\Gamma_{1 0}^{1 1}+\Gamma_{1 1}^{1 0})\right]},
\end{eqnarray}
and with the additional definitions:
\begin{eqnarray}
\label{eq.5.16}
k_{1}^{BCBD}\equiv 2\sqrt{
(\Gamma_{1 0}^{1 1})^{2}+(\Gamma_{1 0}^{0 0})^{2}+4\Gamma_{0 0}^{1 0}\Gamma_{1 1}^{1 0}-2\Gamma_{1 0}^{1 1}\Gamma_{1 0}^{0 0} }>0\;\; ,\text{and}\;\;
k_{2}^{BCBD}\equiv2\left[\Gamma_{1 0}^{1 1}+\Gamma_{1 1}^{1 0}-(\Gamma_{1 0}^{0 0}+\Gamma_{0 0}^{1 0})\right] >0,
\end{eqnarray}
where the reaction-rates appearing in (\ref{eq.5.15})-(\ref{eq.5.16}) are those defined in (\ref{eq.3.3}).

(ii) For the model RDCI, setting
\begin{eqnarray}
\label{eq.5.18}
\phi_{RDCI}\equiv \frac{2\Gamma_{0 0}^{1 0}-\Gamma_{1 0}^{1 1}+\sqrt{(\Gamma_{1 0}^{1 1})^{2}+4\Gamma_{0 0}^{1 0}\Gamma_{0 1}^{1 0}}}{2\left[\Gamma_{0 0}^{1 0}-(\Gamma_{1 0}^{0 1}+\Gamma_{1 0}^{1 1})\right]}\;\; 
\end{eqnarray}
and with the additional definitions:
\begin{eqnarray}
\label{eq.5.19}
k_{1}^{RDCI}\equiv2\sqrt{(\Gamma_{1 0}^{1 1})^{2}+4\Gamma_{0 0}^{1 0}\Gamma_{0 1}^{1 0}}>0\;\; \text{and}\;\;
k_{2}^{RDCI}\equiv 2\left[(\Gamma_{0 1}^{1 0}+\Gamma_{1 0}^{1 1})-\Gamma_{0 0}^{1 0}\right] >0.
\end{eqnarray}
The reaction-rates appearing in (\ref{eq.5.18})-(\ref{eq.5.19}) are those introduced in section VI.

Under the conditions (i) and (ii), at the continuum mean-field level, we have for the model BCBD and RDCI, with  $\widetilde{\rho}_{MF}^{l}(x,t)\equiv\rho_{MF}^{l}(x,t)-\phi_{l} $, where $l=BCBD,RDCI$, equations of motion which 
are {\it Fisher's equations}:
\begin{eqnarray}
\label{eq.5.21}
\frac{\partial}{\partial t}\widetilde{\rho}_{MF}^{l}(x,t)= \frac{k_{2}^{l}}{2}\frac{\partial^{2}}{\partial x^{2}}\widetilde{\rho}_{MF}^{l}(x,t)+k_{1}^{l}\widetilde{\rho}_{MF}^{l}(x,t)-k_{2}^{l} (\widetilde{\rho}_{MF}^{l}(x,t))^{2}
\end{eqnarray}
Assuming $L$ to be even and relabelling the sites of the chain according to the shift: $x\longrightarrow x-\frac{L}{2}$, we consider an initial inhomogeneous configuration with
\begin{eqnarray}
\label{eq.5.22}
\langle n_{x}(0)\rangle=
\left\{
\begin{array}{l l}
\langle n_{x}(\infty)\rangle, \text{if $x\in[-L/2,0]$}\\
0,\;\; \text{otherwise}
\end{array}
\right.
\end{eqnarray}
We want now to compare the prediction of the mean-field theory with the results obtained directly from the microscopic results derived in section V and VI and thus  compute the time-dependent position $X(t)$ of the wave-front and its time-dependent
 width $w(t)$. This is done according to the formulae \cite{Riordan}:
\begin{eqnarray}
\label{eq.5.23}
X(t)&=&\sum_{x=-L/2}^{L/2}\frac{\langle n_{x}(t)\rangle}{\langle n_{x}(\infty)\rangle }-\frac{L}{2},\;\; \text{and}\;\;
w(t)^{2}=2\sum_{x=-L/2}^{L/2}\frac{x\langle n_{x}(t)\rangle}{\langle n_{x}(\infty)\rangle }
-\left(\frac{L}{2}\right)^{2}-X(t)^{2}
\end{eqnarray}
With help of  (\ref{eq.3.6.0}) and   (\ref{eq.4.6.0}), and denoting with  $q_{l},\delta_{l},\gamma_{l}, E_{l}^{\ast}$ the quantities related to the model $l\in(BCBD,RDCI)$ and defined (computed) in sections V and VI, we obtain in the long-time regime $(E^{\ast}_{l}\delta_{l}+\gamma_{l})t\gg 1$ where the time scales as $t\propto L^{2}$:
\begin{eqnarray}
\label{eq.5.24}
X_{l}(t)=\sqrt{2u_{l}(E^{\ast}_{l}\delta_{l}+\gamma_{l})t}\;\left[1+{\cal O}(e^{-\frac{1}{2}\sqrt{L^{2}/u_{l}}})\right]
\end{eqnarray}
and
\begin{eqnarray}
\label{eq.5.25}
w_{l}(t)=\sqrt{u_{l}(E^{\ast}_{l}\delta_{l}+\gamma_{l})t}\;\left[1+{\cal O}(e^{-\frac{1}{2}\sqrt{L^{2}/u_{l}}})\right],
\end{eqnarray}
where we have introduced the parameter $u_{l}\equiv\frac{L^{2}}{2(E^{\ast}_{l}\delta_{l}+\gamma_{l})t}={\cal O}(1)$.

From these exact results, it appears that the location of the wave-front moves as  $\sqrt{t}$. Moreover, in contrast to the {\it Fisher's mean-field theory}, the width of the wave-front broadens as $\sqrt{t}$. These results which have also been observed, in the continuum limit, for the one-dimensional reversible diffusion-coagulation (without input) model \cite{Riordan}, confirms that in one spatial dimension the mean-field Fisher's picture fails. In fact Riordan et {\it al.} \cite{Riordan} have argued on the basis of numerical computations for the  reversible diffusion-coagulation (without input) model that in higher dimension ($\geq 4$) the latter model is in agreement with Fisher's mean-field predictions: the width of the wave-front does not broaden. Very recently other authors who studied the same model as Riordan {\it et al.} (in dimensions $d>1$) came to completely different conclusions \cite{Moro}.

Furthermore, for $\widetilde{\rho}_{MF}^{l}$, Fisher's equation  admits two (homogeneous) stationary-states, namely $\widetilde{\rho}_{MF}^{l}(\infty)=\frac{k_{1}^{l}}{k_{2}^{l}} $ which is linearly stable \cite{Murray} and $\widetilde{\rho}_{MF}^{l}(\infty)=0 $ which is linearly unstable \cite{Murray}. This implies that at mean-field level,  we would have for the stationary density, $\rho_{MF}^{l}(\infty)=\widetilde{\rho}_{MF}^{l}(\infty)+\phi_{l}$  which corresponds to two possible steady-states: $\rho_{MF}^{l}(\infty)= \frac{k_{1}^{l}}{k_{2}^{l}}+\phi_{l},\;\rho_{MF}^{l}(\infty)=\phi_{l}$.
However, from the exact expressions of the stationary density (\ref{eq.3.4}) and (\ref{eq.4.4}), we know that the models under consideration admit unique steady-states which do not coincide with the mean-field prediction.

\section{Summary and conclusion}

In this work we have extended the {\it conventional} interparticle distribution function (IPDF) method. We introduced  a {\it string-function} which is a natural generalization of the {\it empty-interval function} employed in the IPDF method. We derived the (five) constraints for the equations of motion to close (see (\ref{eq.1.3}) and (\ref{eq.1.4})). We  solved  the equation of motion of this string-function on a periodic and finite lattice for the general form of a class of models which cannot be mapped onto free-fermion systems and which so far (to our knowledge) have been poorly understood (see (\ref{eq.2.14})). Then we specifically studied two models: The first one, which  is a model with branching, coagulation, birth and death processes (the BCBD model), can be viewed as a generalization of the voter model and/or as an epidemic model. The BCBD model is an example of model which {\it cannot} be solved directly by the traditional IPDF method. For this model, under certain restrictions on the reaction-rates (see (\ref{eq.3.3})), the density, the non-instantaneous two-point  as well as the exact nearest-neighbour (instantaneous) correlation functions have been analyzed:  the steady-states (see 
(\ref{eq.3.4})-(\ref{eq.3.5})) as well as the dynamical approach towards the latter has been computed exactly (see (\ref{eq.3.7})-(\ref{eq.3.8})). In particular the relaxational spectrum as well as the the {\it inverse of the relaxation-time} have been obtained (see (\ref{eq.3.6.0})). A similar analysis has been performed for a reversible diffusion-coagulation model with input of particles (RDCI model). The latter  (with the {\it usual} restriction that the coagulation rate coincides with the diffusion one) can be studied with help of the {\it traditional} IPDF method (the {\it string-function} then reduces to the {\it empty-interval function}). In addition to the above-mentioned quantities, which we were  able to compute also for the RDCI model, we calculated  the stationary {\it interparticle-function}  (see (\ref{eq.4.12})).

On the basis of the exact results, we have developped an approximative recursive scheme that allows to compute the (other) instantaneous two-point correlation functions (see (\ref{eq.2.17.5})). 

Studying these  (BBCD and RDCI) models, we observed that the latter are characterized by a translationally-invariant stationary distributions,  for which contrary to what happens to free-fermion systems correlations are present: $\langle n_{x}n_{x+1}\rangle(\infty)\neq\langle (n_{x}(\infty)\rangle)^{2} $. 

Later we studied the solution  of the RDCI model and its implications on other systems  related via similarity transformations. In particular we considered a  class of  similarity transformations (see (\ref{eq.5.3}),(\ref{eq.5.4})) which transforms the {\it conventional} empty-interval function into a more general string-function. In so doing we saw that it is possible to map the RDCI model onto the BCBD one , which turns out to be solvable (via the similarity transformation (\ref{eq.5.3}), (\ref{eq.5.4})) with the same constraints encountered in section V. We therefore conclude that the approaches of the section V and VII for solving the BCBD model are {\it equivalent}. However it has to be noticed that working with the generalized {\it string-function} as in section V gives {\it naturally} access to the  solution of the BCBD model without requiring the solution of  another (the RDCI) model.

 We also have identified a model of reversible diffusion-coagulation with particles input and pair annihilation (RDCIPA) which can be mapped, for some choices of the reaction-rates (see (\ref{eq.5.12})-(\ref{eq.5.14})), onto the RDCI model. For this RDCIPA model all the quantities previously computed for the RDCI can be immediately obtained via the similarity transformation (see (\ref{eq.5.8})).

Finally we observed that on some parameter manifold, the mean-field approximation
 of the  BCBD and RDCI models are described (in the continuum limit) by the so-called {\it Fisher equations} which predict that an inhomogeneous initial configuration will evolve without broadening  of the wave-front in the density of particles. Computing the width of the wave-front which broadens as $\sqrt{t}$ (see (\ref{eq.5.25})), we show that the Fisher's mean-field description fails in one-dimension. Another failure of the mean-field theory is observed when one compares the mean-field predictions for the steady-states of the density with the exact results which shows that these stationary states are in fact unique.

\section{Acknowledgments}
We thank E. Moro for pointing out his recent work (Ref.\cite{Moro}).
The support of Swiss National Fonds is gratefully acknowledged.
%


%
\end{document}